\documentclass[reprint, aps, prc, twocolumn, showpacs, showkeys, superscriptaddress, longbibliography, nolinenumbers, floatfix]{revtex4-2}
\usepackage{amsmath}
\usepackage{amssymb}
\usepackage{amsfonts}
\usepackage{graphicx}
\usepackage{float}
\usepackage{dcolumn}
\usepackage{multirow}
\usepackage{natbib}
\usepackage{tcolorbox}
\usepackage[colorlinks = true, linkcolor = blue, urlcolor = blue,  citecolor = blue, anchorcolor = blue]{hyperref}
\usepackage{enumerate}
\usepackage{times}
\usepackage{placeins}

\begin{document}

\title{Structural response of $^{64}$Ge and $^{130}$Cd to superstrong magnetic fields:\\ A covariant density functional study}

\author{I.~Mavroudis}
\affiliation{Department of Theoretical Physics, Aristotle University of Thessaloniki, 54124 Thessaloniki, Greece}

\author{D.~Petrellis}
\affiliation{Department of Theoretical Physics, Aristotle University of Thessaloniki, 54124 Thessaloniki, Greece}

\author{G.~A.~Lalazissis}
\affiliation{Department of Theoretical Physics,	Aristotle University of Thessaloniki, 54124 Thessaloniki, Greece}

\author{P.~Vesel\'y}
\affiliation{Nuclear Physics Institute, Czech Academy of Sciences, CZ-250 68 Re\v{z} near Prague, Czech Republic}

\author{P.~Ring}
\affiliation{Department of Physics, Technische Universit\"at M\"unchen, D-85747 Garching, Germany}

\author{Ch.~C.~Moustakidis}
\affiliation{Department of Theoretical Physics, Aristotle University of Thessaloniki, 54124 Thessaloniki, Greece}

\begin{abstract}
\begin{description}
\item[Background]
Superstrong magnetic fields, potentially present in magnetars or transient astrophysical events, break time-reversal symmetry and may modify both single-particle energy levels and the nuclear structure of finite nuclei. The different magnetic couplings of neutrons and protons allow their spatial distributions to respond differently.
\item[Purpose]
We investigate how the magnetic response of $^{64}$Ge and $^{130}$Cd, which are waiting points in the $rp$- and $r$-process, respectively, relates changes in the bulk observables to modifications in their microscopic structure. We also assess the dependence of this response on the choice of covariant energy density functional.
\item[Methods]
Self-consistent and axially symmetric calculations are performed using the NL3* and DD-ME2 functionals, considering the minimal coupling of protons to an external uniform magnetic field, the anomalous magnetic moments of nucleons, and the induced time-odd fields. Pairing correlations are neglected. The analysis examines the total energy, quadrupole deformation, radii, spatial density distributions, single-particle spectra, and meson-field energy contributions.
\item[Results]
Both functionals predict a decrease in the total energy and an overall increase in nucleus size at high magnetic field values, accompanied by nonmonotonic changes of deformation between prolate and oblate configurations. The neutron skin of $^{130}$Cd and the proton skin of $^{64}$Ge increase overall. The spacing between the highest occupied and lowest unoccupied neutron single-particle levels narrows overall in $^{130}$Cd, while in $^{64}$Ge, it initially narrows and later reopens at higher fields. Also, at high fields, the shifts, relative to zero field, in the scalar and time-like vector energy contributions largely cancel, with their net contribution partially opposing the decrease in total energy. The meson-current energies provide smaller attractive contributions.
\item[Conclusions]
Both nuclei expand at high fields, but the expansion accompanies different changes in their single-particle levels and nucleon densities. NL3* and DD-ME2 agree on these qualitative trends. NL3* predicts a thicker neutron skin in $^{130}$Cd throughout the field range, and the magnitudes of the current energies also depend on the functional. These static calculations motivate further studies of magnetic-field-dependent nucleon separation energies, weak-transition strengths, and nuclear reaction and decay rates to assess their possible implications for nucleosynthesis in strongly magnetized environments.
\end{description}
\end{abstract}



\maketitle

\section{Introduction}
\label{sec:introduction}

Extreme magnetic fields may modify single-particle energy levels and the nuclear structure of finite nuclei \cite{Kondratyev_2018}. Magnetars provide the primary astrophysical motivation, with surface dipolar magnetic fields estimated to be of the order of $10^{14}$-$10^{15}$~G from their observed spin-down rates \cite{Kouveliotou1998,Turolla_2015,Kaspi2017}, and their transient activity includes giant gamma-ray flares associated with rapid magnetic-energy release \cite{Palmer2005}. Relativistic stellar models permit much stronger internal fields, approaching $10^{18}$~G \cite{Bocquet1995,Cardall2001}. These interior-field scales are theoretical estimates, distinct from the inferred exterior fields, but they motivate the examination of the nuclear structure when magnetic interactions become comparable to intrinsic nuclear energy scales.

Dynamical environments provide additional routes to strong fields. Three-dimensional simulations find magnetic amplification and a large-scale dynamo action in rapidly rotating proto-neutron stars formed during stellar core collapse \cite{Mosta2015}, and rapid amplification through the Kelvin-Helmholtz instability during binary-neutron-star mergers \cite{Kiuchi2015}. For relativistic heavy-ion collisions in the laboratory, theoretical calculations predict intense, short-lived electromagnetic fields \cite{DengHuang2012}, while charge-dependent directed-flow measurements provide experimental evidence for their effects on quark-gluon plasma \cite{STAR2024EM}.

The characteristic magnetic energy scale for a nucleon is $\mu_N B\simeq0.315[B/(10^{17}\,\mathrm{G})]$~MeV, where $\mu_N$ is the nuclear magneton. Close to $10^{18}$~G this scale becomes comparable to single-particle level spacings, and the magnetic coupling of nucleon spins and proton orbital motion can compete with the nuclear spin-orbit interaction \cite{Arteagaetal2011,Steinetal2016}. Protons and neutrons respond differently, because their magnetic moments have opposite signs and different magnitudes. In a relativistic framework, proton minimal coupling and the anomalous magnetic moments of both species account for these interactions \cite{Broderick2000,Arteagaetal2011}. The external field breaks the time-reversal symmetry, lifts Kramers degeneracy, and induces currents that feed back into the self-consistent mean fields.

Early shell-correction studies established that strong fields can alter shell oscillations and the relative binding of different nuclei \cite{PhysRevLett.84.1086,Kondratyev_2001,Kondratyev_2002}. Subsequent self-consistent studies have used covariant and Skyrme energy-density-functional approaches. Pe\~na Arteaga \textit{et al.}\ incorporated the orbital coupling of protons and the magnetic-moment couplings of nucleons to an external magnetic field, while treating induced nuclear currents and axial deformation self-consistently within covariant density functional theory (CDFT) \cite{Arteagaetal2011}, using the NL3 parametrization. Their calculations of $^{16}$O and nuclei in the iron region connected field-induced single-particle level rearrangements to changes in energy, radius, and shape. Basilico \textit{et al.}\ extended this treatment to field-dependent nuclear masses, using the NL3 and DD-ME2 parametrizations, and incorporated them into calculations for the equilibrium composition of magnetars' outer crust \cite{PhysRevC.92.035802}. Changes in binding energy and single-particle structure  have also been discussed in RMF-based applications to electron-capture and neutrino-energy-loss rates in strongly magnetized stellar matter \cite{LiuGu2016,LiuLiu2018,Liu2021PASP,Liu2021RAA}.

Within the nonrelativistic framework, Stein \textit{et al.}\ used three-dimensional Skyrme Hartree-Fock calculations with the SV-bas parametrization, for $^{12}$C, $^{16}$O, and $^{20}$Ne \cite{Steinetal2016}. Their explicit orbital and spin couplings produced changes in occupations, magnetization, currents, and deformation. Afterwards, Jiang and Chen used the SLy6 functional in Sky3D to calculate field-dependent properties of 4110 nuclides with $5\leq Z\leq82$ and examine their consequences for the outer crust \cite{Jiang}. Then, Jiang \textit{et al.}\ (2025) compared the UNEDF1 and modified UNEDF1*  parametrizations over $10^{14}$-$10^{18}$~G, investigating how magnetic-field-induced changes in nuclear masses and the resulting outer-crust composition depend on the symmetry energy \cite{Jiang_2025}.

More recently, Jiang \textit{et al.}\ (2026) examined $^{56}$Fe, $^{80}$Zn, and $^{132}$Sn with the SLy4d parametrization, investigating the effects of time-odd contributions and the proton diamagnetic term on nuclear energies, density distributions, deformation, and single-particle structure \cite{Jiangetal2026}. An extension to single-$\Lambda$ hypernuclei and their nuclear cores uses the SLy4 parametrization together with the SLL4 $\Lambda N$ interaction to study how the impurity modifies radii, deformation, level rearrangements, and separation energies \cite{Jiang2026Lambda}.

CDFT provides a relativistic framework for describing the coupling of nucleons to an external magnetic field. Its formulation for finite nuclei is based on scalar and vector mean fields, with nuclear interactions described through nonlinear meson self-interactions or density-dependent meson-nucleon couplings \cite{Walecka1974,BogutaBodmer1977,Gambhir1990,Ring1996}. The Dirac structure naturally generates the nuclear spin-orbit interaction from the scalar and vector mean fields, without requiring a separate phenomenological spin-orbit term \cite{Ring1996,Vretenar2005,Arteagaetal2011}. In meson-exchange CDFT, Lorentz covariance requires a common coupling for the time-like and spatial components of each vector meson. The time-odd mean fields generated by nuclear currents, therefore, do not introduce additional adjustable couplings \cite{PhysRevC.81.014309,Arteagaetal2011}. In Skyrme functionals, local gauge invariance likewise relates several time-even and time-odd couplings \cite{DobaczewskiDudek1995}.

The choice of $^{64}$Ge and $^{130}$Cd is motivated by their roles in different nucleosynthesis processes. In the rapid proton-capture ($rp$) process, $^{64}$Ge is a waiting point, whose effective lifetime depends on the competition between $\beta^+$ decay and proton-capture paths influenced by the masses of neighboring nuclei. Small proton separation energies favor reverse photodisintegration reactions, which hinders net proton capture, while sequential two-proton capture can provide a competing path \cite{Schatz1998,Clark2007,Zhou2023}. The neutron-rich nucleus $^{130}$Cd, with $N=82$, lies in the waiting-point region of the rapid neutron-capture ($r$) process, where nuclear shell structure, neutron separation energies, and $\beta^-$-decay rates influence the reaction pathway and the formation of the $A\simeq130$ abundance peak \cite{Cowan2021,Atanasov2015,Manea2020}. Mass measurements in the vicinity of these nuclei have demonstrated the sensitivity of the corresponding nucleosynthesis pathways to neighboring nuclear masses \cite{Zhou2023,Atanasov2015}. The two nuclei also exhibit distinct structural characteristics, as $^{64}$Ge is an $N=Z$ nucleus with experimentally established quadrupole collectivity \cite{Starosta2007}, while isomer spectroscopy of $^{130}$Cd provides information on neutron-rich $N=82$ shell closure \cite{Jungclaus2007}.

Magnetic fields can influence the evolution of waiting points through changes in both the nuclear structure \cite{Kondratyev_2025} and the surrounding lepton phase space. Calculations of stellar $\beta^+$-decay and electron-capture rates for $rp$-process waiting-point nuclei have shown that magnetic-field-induced modifications of the leptonic phase space can substantially alter weak-interaction rates \cite{wg2d-fx7j}. However, such changes do not necessarily imply a corresponding modification of the nuclear structure. So, investigating the intrinsic magnetic response of $^{64}$Ge and $^{130}$Cd provides a complementary step toward understanding how strong magnetic fields may affect the nuclear inputs to nucleosynthesis calculations.

In the present work, we calculate the static response of $^{64}$Ge and $^{130}$Cd to fields up to $10^{18}$~G, using the NL3* \cite{LALAZISSIS200936} and DD-ME2 \cite{PhysRevC.71.024312}. These parametrizations employ nonlinear meson self-interactions and density-dependent meson-nucleon couplings, respectively. We examine the evolution of total energy, nuclear deformation, radius, and skin thickness in relation to changes in the single-particle spectra and nucleon density distributions. The scalar, time-like vector, and current-energy contributions are analyzed separately to clarify the role of the self-consistent mean fields in the magnetic response of these nuclei.

The present calculations provide a basis for extending the study to neighboring nuclei and obtaining consistent field-dependent mass differences, separation energies, and reaction $Q$ values. Previous studies combining relativistic mean-field (RMF) calculations with the proton-neutron quasiparticle random-phase approximation (pn-QRPA) have demonstrated the sensitivity of stellar weak-decay rates to nuclear deformation in neutron-rich $r$-process waiting-point nuclei, including $^{130}$Cd \cite{Nabietal2021}. Together with corresponding reaction and decay rates, these nuclear inputs could be incorporated into nucleosynthesis networks to assess possible changes in waiting-point lifetimes and abundance distributions. Time-dependent extensions of the present framework could also be used to investigate nuclear excitations and dynamical responses to transient electromagnetic fields \cite{Nakatsukasa2016NuclearDynamics}.

Section~\ref{tf:sec:framework} presents the theoretical framework, Sec.~\ref{sec:results} discusses the results, and Sec.~\ref{sec:conclusions} gives the conclusions. Appendix~\ref{tf:sec:axial} details the axially symmetric formulation, Appendix~\ref{tf:sec:energy} defines the nuclear energy and center-of-mass correction, and Appendix~\ref{app:additional_results} provides the quantum-number assignments for the single-particle spectra. The Supplemental Material~\cite{SupplementalMaterial} contains additional proton and neutron spectra and their quantum-number assignment tables for the remaining nucleus-functional combinations.

\section{Theoretical framework}
\label{tf:sec:framework}

We describe finite nuclei in a prescribed, static magnetic field within covariant density functional theory, using the NL3* \cite{LALAZISSIS200936} and DD-ME2 \cite{PhysRevC.71.024312} parametrizations. The coupling to the magnetic field follows the finite-nucleus framework of Refs.~\cite{Arteagaetal2011,PhysRevC.92.035802}. The nuclear fields are treated in the Hartree and no-sea approximations. Pairing correlations are neglected, following the strong-field CDFT treatment of Pe\~na Arteaga \textit{et al.}\ \cite{Arteagaetal2011}. This approximation is motivated by magnetic suppression of  pairing effects, which can be quenched at sufficiently strong fields \cite{Kittel1963}. Natural units $\hbar=c=1$ and the metric $g_{\mu\nu}=\operatorname{diag}(1,-1,-1,-1)$ are used. Unless otherwise specified, bold symbols denote spatial vectors and arrows denote isospin vectors.

\subsection{Effective Lagrangian and magnetic couplings}
\label{tf:sec:lagrangian}

The degrees of freedom are the nucleon doublet $\psi$, the isoscalar scalar field $\sigma$, the isoscalar vector field $\omega_\mu$, the isovector vector field $\vec\rho_\mu$, and the electromagnetic potential $A_\mu$ generated by the protons. The external potential $A_\mu^{\rm ext}$ is specified independently.
The Lagrangian density is
\begin{equation}
  \mathcal L=\mathcal L_N+\mathcal L_m+\mathcal L_{\rm int}
             +\mathcal L_{BO}+\mathcal L_{BM},
\label{tf:eq:lagrangian}
\end{equation}
where the free-nucleon term is
\begin{equation}
  \mathcal L_N=\bar\psi(i\gamma^\mu\partial_\mu-m)\psi,
\label{tf:eq:free-nucleon}
\end{equation}
with $\bar\psi=\psi^\dagger\gamma^0$. The meson and internal electromagnetic terms are
\begin{equation}
\begin{aligned}
  \mathcal L_m={}&\frac12\partial_\mu\sigma\,\partial^\mu\sigma
                 -\frac12m_\sigma^2\sigma^2-U_{\rm nl}(\sigma)
  \\
  &-\frac14\Omega_{\mu\nu}\Omega^{\mu\nu}
                 +\frac12m_\omega^2\omega_\mu\omega^\mu
  \\
  &-\frac14\vec R_{\mu\nu}\cdot\vec R^{\mu\nu}
                 +\frac12m_\rho^2\vec\rho_\mu\cdot\vec\rho^{\,\mu}
  \\
  &-\frac14F_{\mu\nu}F^{\mu\nu}.
\end{aligned}
\label{tf:eq:meson-sector}
\end{equation}
The interaction with the nucleons reads
\begin{equation}
\begin{aligned}
  \mathcal L_{\rm int}={}&-g_\sigma\bar\psi\sigma\psi
    -g_\omega\bar\psi\gamma^\mu\omega_\mu\psi
  \\
  &-g_\rho\bar\psi\gamma^\mu(\vec\tau\cdot\vec\rho_\mu)\psi
    -e_p\bar\psi\gamma^\mu A_\mu Q_p\psi .
\end{aligned}
\label{tf:eq:interaction}
\end{equation}
The vertices $g_i$ are constant for NL3* and density dependent for DD-ME2. The isospin matrices $\vec\tau$ are Pauli matrices, with $\tau_3\psi_n=+\psi_n$ and $\tau_3\psi_p=-\psi_p$, so that
\begin{equation}
  Q_p=\frac{1-\tau_3}{2},\qquad Q_n=\frac{1+\tau_3}{2}.
\label{tf:eq:projectors}
\end{equation}
Here $e_p>0$ and $e_p^2/(4\pi)\simeq1/137.036$.
There is no additional factor of $1/2$ in the $\rho$-meson vertex. The field tensors are
\begin{equation}
\begin{aligned}
  \Omega_{\mu\nu}&=\partial_\mu\omega_\nu-\partial_\nu\omega_\mu,\\
  \vec R_{\mu\nu}&=\partial_\mu\vec\rho_\nu-\partial_\nu\vec\rho_\mu,\\
  F_{\mu\nu}&=\partial_\mu A_\nu-\partial_\nu A_\mu,\\
  F^{\rm ext}_{\mu\nu}&=\partial_\mu A^{\rm ext}_\nu
                        -\partial_\nu A^{\rm ext}_\mu.
\end{aligned}
\label{tf:eq:tensors}
\end{equation}
Only the third isospin component of $\vec\rho_\mu$ survives in the charge-conserving Hartree solution, and its tensor therefore has the Abelian form used above.

On the nuclear length scale, the external field is taken to be uniform and directed along the $z$ axis. With $A^\mu=(A^0,\mathbf A)$, we choose the symmetric gauge
\begin{equation}
\begin{aligned}
  \mathbf B&=B\hat{\mathbf z},\qquad A^{0}_{\rm ext}=0,\\
  \mathbf A^{\rm ext}&=\frac12\mathbf B\times\mathbf r
    =\frac B2(-y,x,0),
\end{aligned}
\label{tf:eq:gauge}
\end{equation}
which gives $\boldsymbol\nabla\times\mathbf A^{\rm ext} =B\hat{\mathbf z}$. The proton charge coupling is
\begin{equation}
  \mathcal L_{BO}=-e_p\bar\psi\gamma^\mu A^{\rm ext}_\mu Q_p\psi.
\label{tf:eq:orbital}
\end{equation}
The additional coupling to the nucleon magnetic moments is
\begin{equation}
\begin{aligned}
  \mathcal L_{BM}&=-\bar\psi\chi^{\rm ext}\psi,\\
  \chi^{\rm ext}&=\frac{\mu_N}{2}\widehat\kappa\,
                  \sigma^{\mu\nu}F^{\rm ext}_{\mu\nu},\\
  \sigma^{\mu\nu}&=\frac{i}{2}[\gamma^\mu,\gamma^\nu],\qquad
  \widehat\kappa=\kappa_pQ_p+\kappa_nQ_n,
\end{aligned}
\label{tf:eq:pauli}
\end{equation}
where $\mu_N=e_p/(2m)$, $\kappa_p=1.7928$, and $\kappa_n=-1.91315$ \cite{Arteagaetal2011,PhysRevC.92.035802}. The minimal Dirac coupling already contributes $\mu_N$ to the proton's magnetic moment, giving a total moment of $(1+\kappa_p)\mu_N$.
For the field in Eq.~\eqref{tf:eq:gauge},
\begin{equation}
  \frac12\sigma^{\mu\nu}F^{\rm ext}_{\mu\nu}=-\Sigma_z B,
  \qquad \chi^{\rm ext}=-\widehat\kappa\mu_N\Sigma_z B,
\label{tf:eq:pauli-reduction}
\end{equation}
with $\boldsymbol\Sigma=\operatorname{diag}(\boldsymbol\sigma_{\rm spin}, \boldsymbol\sigma_{\rm spin})$.

The external potential is kept fixed in the variational problem. The Maxwell term in Eq.~\eqref{tf:eq:meson-sector} contains only the internal field, and the energy of the prescribed external field is excluded from the nuclear energy. As in Refs.~\cite{Arteagaetal2011,PhysRevC.92.035802}, the Pauli term couples to $F^{\rm ext}_{\mu\nu}$, while the internal electromagnetic field retains its minimal coupling.

\subsection{NL3* and DD-ME2 parametrizations}
\label{tf:sec:parametrizations}

NL3* employs constant meson-nucleon vertices and the nonlinear scalar potential \cite{LALAZISSIS200936}
\begin{equation}
  U_{\rm nl}(\sigma)=\frac{g_2}{3}\sigma^3+
                     \frac{g_3}{4}\sigma^4.
\label{tf:eq:nonlinear}
\end{equation}
The fitted coefficients are $g_2=-10.8093\,\mathrm{fm}^{-1}$ and $g_3=-30.1486$, used with $-U_{\rm nl}$ in the Lagrangian. The quadratic scalar mass term is already included separately in Eq.~\eqref{tf:eq:meson-sector}.

For DD-ME2, $U_{\rm nl}=0$ and $g_2=g_3=0$, while the medium dependence instead enters through the vertices \cite{PhysRevC.71.024312,Niksic2002}. Their argument is the invariant baryon density
\begin{equation}
  \rho_v=\sqrt{j_\mu j^\mu},\qquad
  x=\frac{\rho_v}{\rho_{\rm sat}},\qquad
  \rho_{\rm sat}=0.152\,\mathrm{fm}^{-3},
\label{tf:eq:density}
\end{equation}
where $j^\mu$ is the baryon current. In the presence of a spatial current, $\rho_v=\sqrt{(j^0)^2-|\mathbf j|^2}$. The density-dependent vertices are
\begin{equation}
\begin{aligned}
  g_i(\rho_v)&=g_i(\rho_{\rm sat})f_i(x),\qquad i=\sigma,\omega,\\
  f_i(x)&=a_i\frac{1+b_i(x+d_i)^2}{1+c_i(x+d_i)^2},\\
  g_\rho(\rho_v)&=g_\rho(\rho_{\rm sat})\exp[-a_\rho(x-1)].
\end{aligned}
\label{tf:eq:dd-vertices}
\end{equation}
Normalization $f_i(1)=1$ maintains the precision of the published parameters. In the more general notation of Ref.~\cite{PhysRevC.89.054320}, $e_i=d_i$ for DD-ME2.

\begin{table}[!t]
\caption{Parameters of NL3* and DD-ME2 are taken from Table~I of Ref.~\cite{PhysRevC.89.054320}, consistent with the original parametrizations \cite{LALAZISSIS200936,PhysRevC.71.024312}. The $g_i$ entries are constant NL3* vertices or DD-ME2 vertices at $\rho_{\rm sat}$. The zero-field symmetric-matter saturation densities $\rho_0$ are taken from Table~V of Ref.~\cite{PhysRevC.89.054320}. A dash denotes an unused parameter and unit $1$ denotes a dimensionless quantity.}
\label{tf:tab:parameters}
\centering
\begin{ruledtabular}
\begin{tabular}{llrr}
Parameter & Unit & NL3* & DD-ME2 \\
\colrule
$m$        & MeV & 939       & 939       \\
$m_\sigma$ & MeV & 502.5742  & 550.1238  \\
$m_\omega$ & MeV & 782.600   & 783.000   \\
$m_\rho$   & MeV & 763.000   & 763.000   \\
$g_\sigma$ & $1$ & 10.0944   & 10.5396   \\
$g_\omega$ & $1$ & 12.8065   & 13.0189   \\
$g_\rho$   & $1$ & 4.5748    & 3.6836    \\
$g_2$ & $\mathrm{fm}^{-1}$ & $-10.8093$ & $0$ \\
$g_3$ & $1$ & $-30.1486$ & $0$ \\
$a_\sigma$ & $1$ & --- & 1.3881 \\
$b_\sigma$ & $1$ & --- & 1.0943 \\
$c_\sigma$ & $1$ & --- & 1.7057 \\
$d_\sigma$ & $1$ & --- & 0.4421 \\
$a_\omega$ & $1$ & --- & 1.3892 \\
$b_\omega$ & $1$ & --- & 0.9240 \\
$c_\omega$ & $1$ & --- & 1.4620 \\
$d_\omega$ & $1$ & --- & 0.4775 \\
$a_\rho$   & $1$ & --- & 0.5647 \\
$\rho_{\rm sat}$ & $\mathrm{fm}^{-3}$ & --- & 0.152 \\
$\rho_0$        & $\mathrm{fm}^{-3}$ & 0.150 & 0.152
\end{tabular}
\end{ruledtabular}
\end{table}

All parameters in Table~\ref{tf:tab:parameters} are retained at finite $B$, without a magnetic-field-dependent refit. The table uses published values. The nonlinear parametrization used here is NL3*, distinct from NL3 employed in Ref.~\cite{Arteagaetal2011}. NL3* has $\rho_0=0.150\,\mathrm{fm}^{-3}$ but requires no density normalization of its constant vertices. For DD-ME2, $\rho_{\rm sat}=0.152\,\mathrm{fm}^{-3}$ remains the fixed reference density in both the vertices and their derivatives, even when the local density changes with $B$.

\subsection{Single-nucleon Dirac equation and rearrangement self-energy}
\label{tf:sec:dirac}

The scalar and vector potentials are defined by
\begin{equation}
\begin{aligned}
  S&=g_\sigma\sigma,\\
  V_\mu&=g_\omega\omega_\mu+g_\rho\tau_3\rho_{3\mu}
          +e_pQ_p(A_\mu+A^{\rm ext}_\mu).
\end{aligned}
\label{tf:eq:potentials}
\end{equation}
The effective Dirac mass is $M=m+S$. Variation with respect to $\bar\psi$ yields
\begin{equation}
  \bigl[\gamma^\mu(i\partial_\mu-V_\mu-\Sigma_{R\mu})
        -M-\chi^{\rm ext}\bigr]\psi=0.
\label{tf:eq:dirac}
\end{equation}
The no-sea sources are sums over occupied positive-energy states,
\begin{equation}
\begin{aligned}
  \rho_s&=\sum_{k\in{\rm occ}}\bar\psi_k\psi_k,&
  j^\mu&=\sum_{k\in{\rm occ}}\bar\psi_k\gamma^\mu\psi_k,\\
  j_3^\mu&=\sum_{k\in{\rm occ}}\bar\psi_k\gamma^\mu\tau_3\psi_k,&
  j_p^\mu&=\sum_{k\in{\rm occ}}\bar\psi_k\gamma^\mu Q_p\psi_k.
\end{aligned}
\label{tf:eq:sources}
\end{equation}
Thus $j_3^\mu=j_n^\mu-j_p^\mu$ in our isospin convention.

For DD-ME2, the density dependence generates the rearrangement self-energy \cite{Niksic2002}
\begin{equation}
  \Sigma_R^\mu=\frac{j^\mu}{\rho_v}
  \bigl[g_\sigma'\sigma\rho_s+g_\omega'\omega_\nu j^\nu
                         +g_\rho'\rho_{3\nu}j_3^\nu\bigr],
\label{tf:eq:rearrangement}
\end{equation}
where $g_i'=\partial g_i/\partial\rho_v$. In particular,
\begin{equation}
\begin{aligned}
  g_i'&=\frac{g_i(\rho_{\rm sat})}{\rho_{\rm sat}}
    \frac{2a_i(b_i-c_i)(x+d_i)}{[1+c_i(x+d_i)^2]^2},
    \quad i=\sigma,\omega,\\
  g_\rho'&=-\frac{a_\rho}{\rho_{\rm sat}}g_\rho(\rho_v).
\end{aligned}
\label{tf:eq:derivatives}
\end{equation}
The scalar contribution in Eq.~\eqref{tf:eq:rearrangement} has a plus sign for the interaction convention in Eq.~\eqref{tf:eq:interaction}. The Lorentz contractions retain the currents, e.g., $\omega_\nu j^\nu=\omega^0j^0-\boldsymbol\omega\cdot\mathbf j$.
In the local baryon rest frame, $\boldsymbol\Sigma_R=0$. For NL3*, all $g_i'$ and $\Sigma_R^\mu$ vanish. The fixed magnetic moments generate no additional rearrangement term.

For stationary states, Eq.~\eqref{tf:eq:dirac} becomes
\begin{equation}
\begin{aligned}
  h_D\psi_k&=\epsilon_k\psi_k,\\
  h_D&=\boldsymbol\alpha\cdot
       (-i\boldsymbol\nabla-\mathbf V-\boldsymbol\Sigma_R)
       +V^0+\Sigma_R^0\\
     &\quad+\beta M-\beta\widehat\kappa\mu_N\Sigma_zB,
\end{aligned}
\label{tf:eq:hamiltonian}
\end{equation}
where $\boldsymbol\alpha=\gamma^0\boldsymbol\gamma$ and $\beta=\gamma^0$. Both the spatial vector potentials and the Pauli term are retained.

\subsection{Self-consistent meson and electromagnetic fields}
\label{tf:sec:fields}

For static configurations, variation of the scalar field gives the Klein-Gordon equation
\begin{equation}
  (-\nabla^2+m_\sigma^2)\sigma
    =-g_\sigma\rho_s-\frac{dU_{\rm nl}}{d\sigma}.
\label{tf:eq:scalar}
\end{equation}
Here $dU_{\rm nl}/d\sigma=g_2\sigma^2+g_3\sigma^3$ for NL3* and vanishes for DD-ME2. The vector meson and internal electromagnetic equations are
\begin{equation}
\begin{aligned}
  \partial_\nu\Omega^{\nu\mu}+m_\omega^2\omega^\mu
    &=g_\omega j^\mu,\\
  \partial_\nu R_3^{\nu\mu}+m_\rho^2\rho_3^\mu
    &=g_\rho j_3^\mu,\\
  \partial_\nu F^{\nu\mu}&=e_pj_p^\mu.
\end{aligned}
\label{tf:eq:vector-fields}
\end{equation}
The vertices on the right-hand sides are evaluated at the local $\rho_v$ for DD-ME2. The internal Maxwell equation has no Pauli source because Eq.~\eqref{tf:eq:pauli} contains only the external field tensor. The external field changes the meson sources through the nucleon states and currents.

Appendix~\ref{tf:sec:axial} gives the axial and reflection-symmetric formulation. Appendix~\ref{tf:sec:energy} defines the nuclear energy and center-of-mass correction.

\section{Results and Discussion}
\label{sec:results}

We examine the ground-state properties of $^{64}$Ge and $^{130}$Cd in the prescribed external magnetic field $\mathbf B=B\hat{\mathbf z}$, comparing NL3* and DD-ME2 functionals within the framework of Sec.~\ref{tf:sec:framework}. The calculations use 20 fermion and 20 boson oscillator shells, retain axial and reflection symmetry, and omit pairing. The predictions cover 76 field values between $0$~G and $10^{18}$~G, for each nucleus and functional. The single-particle energy spectra are shown at intervals of $10^{17}$~G over the same range, while the spatial densities are compared at $0$~G and $7\times10^{17}$~G. Paired numerical results are given below in order $(NL3^*,\ DD-ME2)$, unless otherwise stated.

\subsection{Bulk properties}
\label{subsec:bulk}

\begin{figure}[!t]
\centering
\includegraphics[width=\columnwidth]{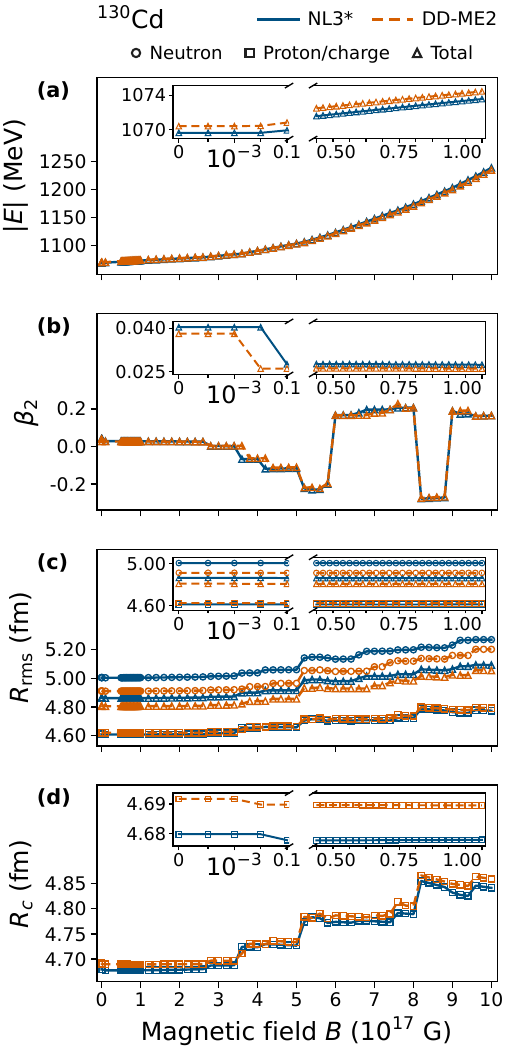}
\caption{Bulk properties of $^{130}$Cd: (a) total-energy magnitude $|E|\equiv|E_{\rm tot}|$, (b) total quadrupole deformation $\beta_2$, (c) neutron, proton, and total rms radii ($R_n$, $R_p$, $R_{\rm m}$), and (d) charge radius $R_c$. Blue solid/orange dashed curves denote NL3*/DD-ME2 in each panel. Circles, squares, and triangles identify neutron, proton/charge, and total quantities, respectively. The main field axes are linear. Each inset uses a symmetric-logarithmic scale up to $0.1$, followed by a linear segment from $0.5$ to $1$. All field values are in $10^{17}$~G.}
\label{fig1}
\end{figure}

\begin{figure}[!t]
\centering
\includegraphics[width=\columnwidth]{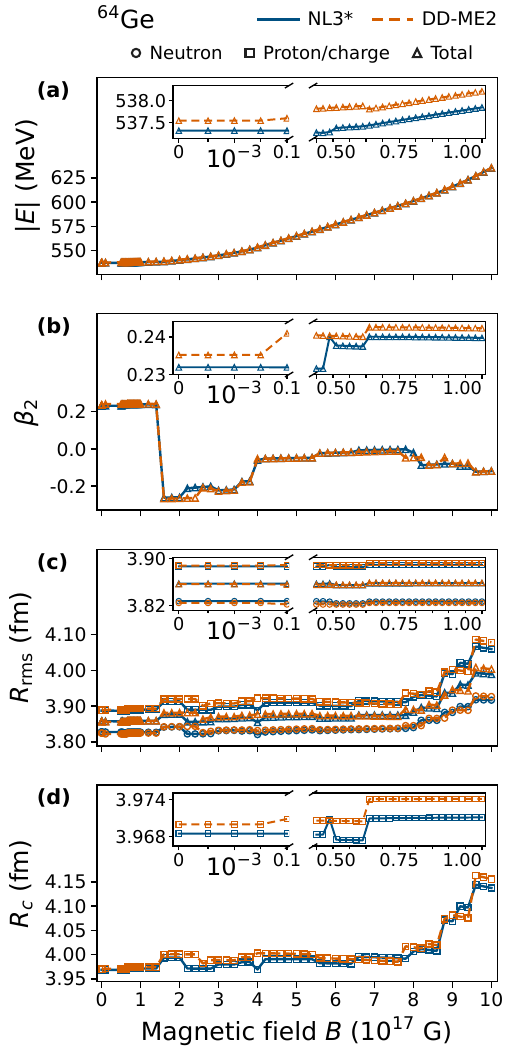}
\caption{Bulk properties of $^{64}$Ge, with panel definitions and conventions as in Fig.~\ref{fig1}.}
\label{fig2}
\end{figure}

Figures \ref{fig1} and \ref{fig2} display the total-energy magnitude ($|E_{\rm tot}|$), quadrupole deformation ($\beta_2$), neutron ($R_n$) and proton ($R_p$) rms radii, total matter ($R_m$) rms radius, and charge radius ($R_c$). The reported $E_{\rm tot}$ is measured relative to the nucleon rest masses. Since, $E_{\rm tot}<0$ throughout the sampled range, a rising curve in panel (a) of Figs. \ref{fig1} and \ref{fig2} represents a decrease in nuclear energy. Radii satisfy $R_{\rm m}^2=(NR_n^2+ZR_p^2)/A$, with $A=N+Z$, and the charge-radius definition is $R_c^2=R_p^2+0.64\,\mathrm{fm}^2$. Thus, the charge-radius curve reflects the proton-radius response with a fixed size correction. The total deformation is $\beta_2=\sqrt{5\pi}\,Q_{20}/(3AR_0^2)$, where $Q_{20}=\int d^3r\,(2z^2-r_\perp^2)(\rho_n+\rho_p)$, $\rho_n$ and $\rho_p$ denote the neutron and proton number densities,  respectively, and $R_0=1.2A^{1/3}$~fm. The positive and negative values of $\beta_2$ denote prolate and oblate quadrupole moments, respectively.

At $B=0$~G, $^{130}$Cd, being a semi-magic nucleus, is nearly spherical, with $\beta_2=(0.040,\ 0.038)$. Its total energy is $(-1069.58,\ -1070.37)$~MeV and its total rms radii are $(4.864,\ 4.808)$~fm. In contrast, $^{64}$Ge has a prolate deformation, with $\beta_2=(0.232,\ 0.235)$, and almost identical, for the two functionals, total rms radii of $3.857$~fm, while its total energy is $(-537.31,\ -537.54)$~MeV. Therefore, the two functionals predict similar zero-field energies for each nucleus.

At higher fields, the energy decrease grows much more rapidly. Between $0$~G and $10^{18}$~G, it reaches $(168.9,\ 164.4)$~MeV for $^{130}$Cd, and $(97.85,\ 97.70)$~MeV for $^{64}$Ge. The corresponding decreases per nucleon are approximately $(1.30,\ 1.26)$~MeV and $(1.53,\ 1.53)$~MeV, for  $^{130}$Cd and $^{64}$Ge, respectively. Both nuclei also expand overall, their total rms radii increase by $(4.72,\ 5.15)\%$ and $(3.40,\ 3.80)\%$, for  $^{130}$Cd and $^{64}$Ge, respectively. Thus, energy decrease does not imply spatial contraction.

The deformation evolves through a sequence of substantially different configurations. For $^{130}$Cd, the initially small positive $\beta_2$ approaches zero near $2.8\times10^{17}$~G, and oblate configurations follow. With both functionals, the configuration changes from $\beta_2\simeq-0.20$ to $(0.166,\ 0.165)$ between $5.8$ and $6.0\times10^{17}$~G. The prolate sequence is interrupted between $8.0$ and $8.2\times10^{17}$~G by a change from approximately $0.205$ to $-0.277$, before prolate configurations reappear between $8.8$ and $9.0\times10^{17}$~G. The functionals agree on this sequence, although the fields in which particular configurations first appear can differ. For example, the first clearly oblate configuration appears in $3.6\times10^{17}$~G with NL3*, and $3.8\times10^{17}$~G with DD-ME2.

For $^{64}$Ge, the configurations change from prolate $\beta_2\simeq0.24$ to oblate $\beta_2\simeq-0.26$ between $1.4$ and $1.6\times10^{17}$~G with both functionals. Subsequently, the oblate deformation decreases in magnitude, reaching $\beta_2=(-0.006,\ -0.014)$ at $7\times10^{17}$~G. Its magnitude then increases again, with the deformation reaching about $-0.119$ at $10^{18}$~G. The occupied orbitals and their self-consistent fields determine the energetically preferred quadrupole deformation.

\subsection{Neutron and proton skins}
\label{subsec:skin}

\begin{figure*}[t]
\centering
\includegraphics[width=\textwidth]{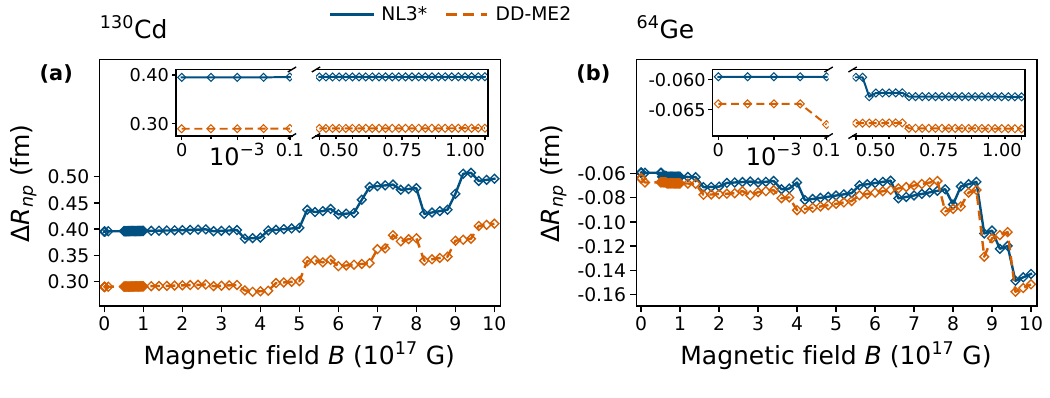}
\caption{Neutron-proton rms-radius difference, $(\Delta R_{np}=R_n-R_p)$, for (a) $^{130}$Cd and (b) $^{64}$Ge.  Functional colors, line styles, and main/inset field-axis conventions as in Fig.~\ref{fig1}.}
\label{fig3}
\end{figure*}

The neutron-proton radius difference, $\Delta R_{np}=R_n-R_p$, in Fig.~\ref{fig3}, separates the relative response of the two species from the overall expansion. A positive value denotes a neutron skin, and a negative value a proton skin. For a deformed nucleus, this single number does not specify a direction-dependent surface thickness.

For $^{130}$Cd, NL3* predicts a thicker neutron skin than DD-ME2, throughout the entire field range. At zero field, the values are $(0.395,\ 0.290)$~fm, while at $10^{18}$~G, they are $(0.496,\ 0.410)$~fm. The increase comes from the larger neutron expansion, because $R_n$ grows by $(0.265,\ 0.290)$~fm, compared to $(0.164,\ 0.170)$~fm for $R_p$. This general trend contains pronounced reversals. Between $8.0$ and $8.2\times10^{17}$~G, for example, the skin decreases by $(0.049,\ 0.043)$~fm, as the configuration becomes oblate. Both radii increase at this switch, but the proton radius increases more. Consequently, an increase in nuclear size does not necessarily imply an increase in neutron-skin thickness.

Although $^{64}$Ge has $N=Z$, its neutron and proton distributions are not identical. The Coulomb interaction and the distinct magnetic couplings break the neutron-proton equivalence. At zero field, its proton skin is $(0.059,\ 0.064)$~fm and grows overall to $(0.143,\ 0.151)$~fm at $10^{18}$~G. Here, the proton radius increases by $(0.172,\ 0.190)$~fm, approximately twice the neutron radius increase of $(0.089,\ 0.102)$~fm. The greatest increases in the proton skin at high fields accompany proton radius steps, particularly near $8.8$ and $9.6\times10^{17}$~G. Between these changes, the skin can decrease even as the energy continues to decrease. The two nuclei thus share an overall expansion but differ in which species expands more. The functional dependence is especially visible in the $^{130}$Cd neutron skin and the relative field-induced changes are similar.

\subsection{Single-particle spectra and occupation rearrangement}
\label{subsec:levels}

\begin{figure*}[t]
\centering
\includegraphics[width=\textwidth]{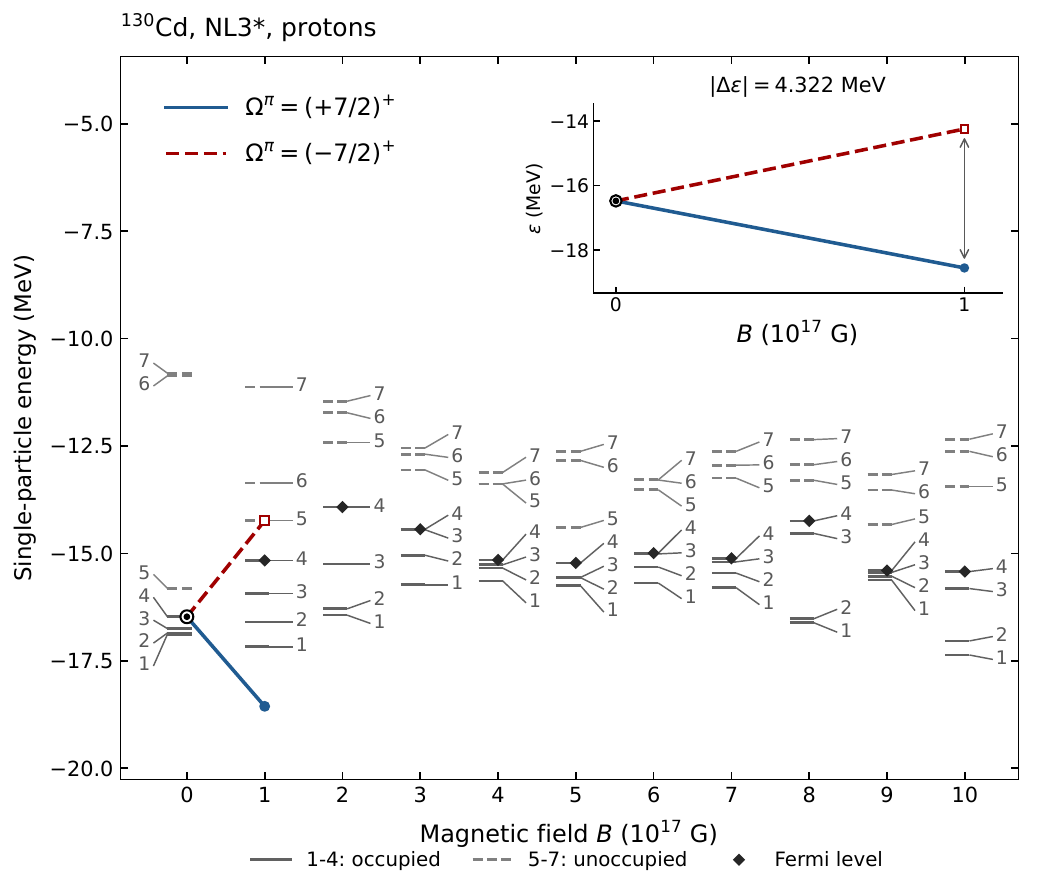}
\caption{Proton single-particle energies of $^{130}$Cd with NL3*. Solid/dashed gray levels denote occupied ranks 1-4/unoccupied ranks 5-7. Ranks 4 and 5 are the highest occupied and lowest unoccupied levels, respectively. Ranks are reassigned at each field. Colored segments connect the $\Omega^\pi=\pm7/2^+$ pair at zero field to its full-spinor overlap matches at $10^{17}$~G. Blue circles/red squares identify positive-/negative-$\Omega$ branches. Filled/open markers denote occupied/unoccupied states, and the black zero-field symbol denotes the degenerate pair. Black diamonds mark the Fermi level at each field. The inset illustrates the $4.322$~MeV splitting. All energy values exclude the nucleon rest mass ($m$), and the inset label $\epsilon$ denotes $\epsilon_k-m$. Assignments can be found in Table~\ref{tab:spe:Cd130:NL3STAR:proton}.}
\label{fig4}
\end{figure*}

\begin{figure*}[t]
\centering
\includegraphics[width=\textwidth]{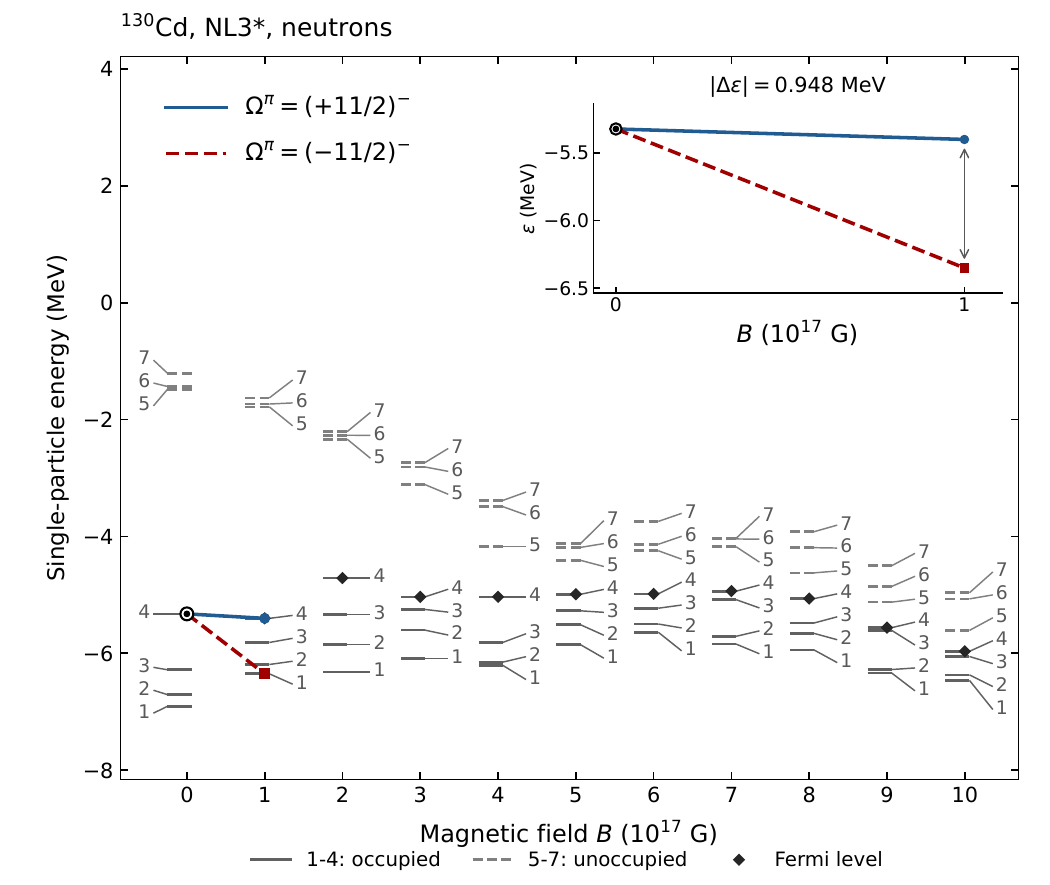}
\caption{Neutron single-particle energies of $^{130}$Cd with NL3*. Colored segments connect the $\Omega^\pi=\pm11/2^-$ pair at zero field to its full-spinor overlap matches at $10^{17}$~G. The inset illustrates the $0.948$~MeV splitting. Plotting and energy conventions as in Fig.~\ref{fig4}. Assignments can be found in Table~\ref{tab:spe:Cd130:NL3STAR:neutron}.}
\label{fig5}
\end{figure*}

Figures~\ref{fig4} and \ref{fig5} illustrate the single-particle proton and neutron spectra near the Fermi level for $^{130}$Cd with NL3*. The remaining nucleus-functional combinations are shown in Figs.~S1-S6 of the Supplemental Material~\cite{SupplementalMaterial}. The plotted single-particle energy levels are $\epsilon_k-m$, with $\epsilon_k$ defined in Eq.~\eqref{tf:eq:hamiltonian} and $m$ the nucleon rest mass. The inset label $\epsilon$ of Figs. \ref{fig4} and \ref{fig5} denotes this rest-mass-subtracted energy. Ranks 1-4 denote the four highest occupied levels and ranks 5-7 the three lowest unoccupied levels, each ordered in energy at the specified field. At zero field, a level represents a degenerate Kramers pair. Thus, neither equal ranks at different fields nor the number of displayed lines identifies a fixed set of orbitals. The assignments in Tables~\ref{tab:spe:Cd130:NL3STAR:proton} and \ref{tab:spe:Cd130:NL3STAR:neutron}, together with Tables~S1-S6 in \cite{SupplementalMaterial}, provide the necessary information. As explained in Appendix~\ref{app:additional_results}, $\Omega$ and parity are good quantum numbers, whereas the oscillator labels and $\Sigma$ describe only the dominant upper-spinor component. In particular, the tabulated $\Sigma$ is not a spin expectation value.

A more specific example of orbital identity is provided by the colored overlays, which match the full Dirac spinors between $B=0$ and $10^{17}$~G within the appropriate signed-$\Omega^\pi$ sectors. For the highest occupied proton pair of $^{130}$Cd, $\Omega^\pi=\pm7/2^+$, NL3* gives a splitting of $4.322$~MeV at the endpoint $10^{17}$~G. The positive-$\Omega$ descendant remains occupied and the negative-$\Omega$ descendant is unoccupied. This is direct evidence for occupation redistribution beyond a simple change of rank. DD-ME2 yields the same occupation pattern with a splitting of $4.491$~MeV (Fig.~S1). The neutron pair $\Omega^\pi=\pm11/2^-$ instead remains fully occupied, with splittings of $0.948$ and $0.870$~MeV for NL3* and DD-ME2, respectively (Fig.~\ref{fig5} and Fig.~S2).

The larger proton splitting is consistent with the additional orbital coupling and the different nucleon magnetic moments in Eq.~\eqref{tf:eq:hamiltonian}. The ordering of the highlighted neutron branches is the reverse of that of the proton branches, as expected for these predominantly spin-aligned components with opposite magnetic-moment signs.

The $^{64}$Ge spectra clarify why splitting and occupation change must be distinguished. The highlighted $\Omega^\pi=\pm7/2^-$ pairs (Figs.~S3-S6) lie below the highest occupied level at zero field. Their proton splittings at $10^{17}$~G are $(4.120,\ 4.184)$~MeV, comparable to those of $^{130}$Cd, while their neutron splittings are $(1.137,\ 1.175)$~MeV. Both descendants remain occupied in each case. A large splitting of a filled pair, therefore, does not, by itself, imply an occupation rearrangement.

A useful local spectral measure is the spacing between the lowest unoccupied and highest occupied displayed levels, $\delta_q(B)=\epsilon_{5q}(B)-\epsilon_{4q}(B)$ where $q=n,p$. Here, subscripts 4 and 5 are displayed ranks. For $^{130}$Cd, the zero-field neutron spacing is $(3.842,\ 4.112)$~MeV, considerably larger than the proton spacing of $(0.662,\ 0.570)$~MeV. The neutron spacing drops to $(0.581,\ 0.770)$~MeV at $5\times10^{17}$~G and to $(0.358, 0.402)$~MeV at $10^{18}$~G, with intermediate nonmonotonic variations. Thus, the local spectral isolation at $N=82$ is substantially reduced in the selected finite-field configurations. The proton spacing fluctuates and reaches $(1.982,\ 1.973)$~MeV at the upper endpoint. For $^{64}$Ge, the neutron spacing instead first decreases from $(1.367,\ 1.295)$~MeV to $(0.477, 0.456)$~MeV at $2\times10^{17}$~G, then reopens nonmonotonically to $(2.501, 2.600)$~MeV at $10^{18}$~G, while the proton spacing at that endpoint is $(0.608,\ 0.685)$~MeV. Therefore, magnetic restructuring does not produce a universal compression of either spectrum.

Changes of the occupied wave functions alter the densities and currents in Eq.~\eqref{tf:eq:sources}, and hence feed back into the proton and neutron Hamiltonians. This provides a physical connection between spectral restructuring and the bulk changes discussed above, consistent with the self-consistent response found in Ref.~\cite{Arteagaetal2011}.

\subsection{Scalar \texorpdfstring{$\sigma$}{σ} and time-like \texorpdfstring{$\omega$}{ω} contributions to the energy}
\label{subsec:mesonfields}

\begin{figure*}[t]
\centering
\includegraphics[width=\textwidth]{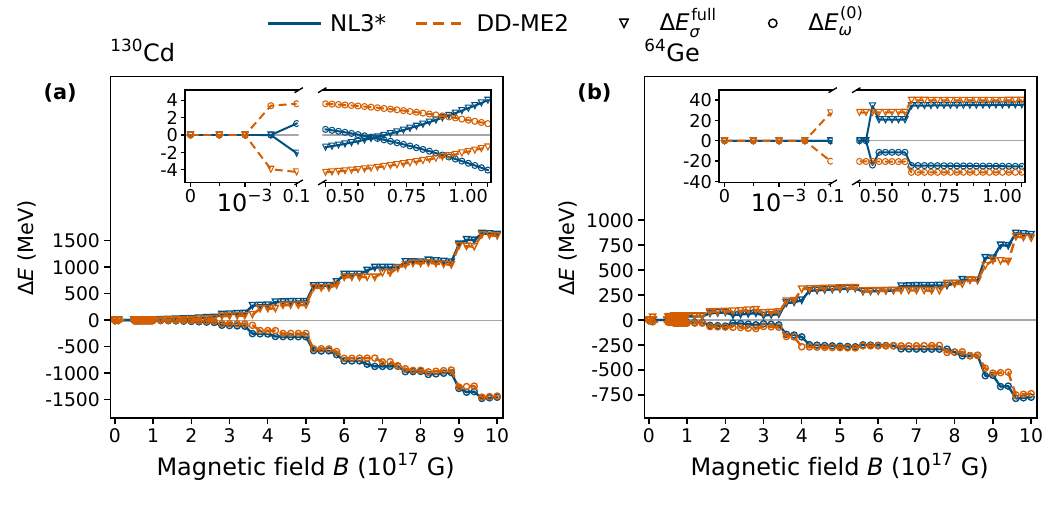}
\caption{Full $\sigma$-channel and time-like $\omega$ energy shifts from $B=0$ for (a) $^{130}$Cd and (b) $^{64}$Ge. The scalar energy includes the bilinear and nonlinear contributions, while the latter vanishes for DD-ME2. The $\omega$ energy excludes its spatial-current contribution. Open downward triangles and circles denote $\Delta E_\sigma^{\mathrm{full}}$ and $\Delta E_\omega^{(0)}$, respectively. Functional colors, line styles, and main/inset field-axis conventions as in Fig.~\ref{fig1}.}
\label{fig6}
\end{figure*}

The energy decomposition in Fig.~\ref{fig6} shows the scalar attraction and time-like vector repulsion as the magnetic field increases. The plotted shifts are $\Delta E_i(B)=E_i(B)-E_i(0)$ for each nucleus and functional, separately. With the energy functional defined in the Appendix~\ref{tf:sec:energy}, the two components are
\begin{equation}
\begin{aligned}
	E_\sigma^{\rm full}&=\frac12\int d^3r\,g_\sigma\sigma\rho_s
	+\int d^3r\,[U_{\rm nl}-\tfrac12\sigma U_{\rm nl}'],\\
	E_\omega^{(0)}&=\frac12\int d^3r\,g_\omega\omega^0 j^0 .
\end{aligned}
\label{tf:eq:energy_shifts}
\end{equation}
The nonlinear correction is $-g_2\sigma^3/6-g_3\sigma^4/4$ under the integral for NL3* and vanishes for DD-ME2. It is included in the plotted scalar energy. Here, the $\omega$ term excludes the spatial-current energy.

At high fields, $\Delta E_\sigma^{\rm full}>0$ and $\Delta E_\omega^{(0)}<0$, so the scalar attraction and time-like $\omega$ repulsion both decrease in magnitude. At $10^{18}$~G, the shifts, expressed as ordered pairs $\{\Delta E_\sigma^{\rm full},\ \Delta E_\omega^{(0)}\}$, are approximately $\{+1616,\ -1455\}$~MeV for $^{130}$Cd with NL3*, and $\{+1590,\ -1434\}$~MeV with DD-ME2. For $^{64}$Ge, they are $\{+856,\ -774\}$~MeV and $\{+821,\ -738\}$~MeV, respectively. These large opposing changes are closely correlated across the field range, including steps associated with changes in deformation and radius. The smaller steps and plateaus are not identical in the two functionals.

The two shifts cancel largely, but their sum is positive: $\Delta E_\sigma^{\rm full}+\Delta E_\omega^{(0)}$ is about $(161,\ 156)$~MeV for $^{130}$Cd, and $(82,\ 83)$~MeV for $^{64}$Ge, at $10^{18}$~G. These two channels therefore oppose, rather than explain, on their own, the net total-energy decrease. The energy adds the free Dirac kinetic term, direct external couplings, isovector and electromagnetic terms, spatial-current contributions, and the center-of-mass correction. In parallel, at $10^{18}$~G, the sum of the external proton minimal-coupling and nucleon Pauli energies is about $(-364,\ -363)$~MeV for $^{130}$Cd, and $(-204,\ -204)$~MeV for $^{64}$Ge. While, the kinetic-energy changes are positive, approximately $(63,\ 67)$~MeV and $(32,\ 32)$~MeV, respectively. The total decrease results from this sum, among competing contributions, not from an increase in the scalar attraction. For DD-ME2, the rearrangement self-energy is essential in the single-nucleon Hamiltonian, but is not an additional independent term in the density-functional energy.

The simultaneous weakening of the two dominant meson energies is compatible with the overall spatial expansion, but it cannot be inferred from the rms radius or central density alone. The scalar source is $\rho_s$, and the time-like $\omega$ source contains $j^0$, while the local vertices also vary with $\rho_v$ in DD-ME2. Shape changes, redistribution within the nucleus, and changes of the Dirac spinors all enter the integrals. Consequently, the total energies can conceal appreciable differences in individual components. A numerical decomposition of a self-consistent solution also does not measure the effect of removing a coupling, which would require recalculating the remaining fields and occupations.

\subsection{Time-reversal breaking and current-energy contributions}
\label{subsec:currents}

The lifting of Kramers degeneracy permits nonvanishing azimuthal currents, which generate the spatial vector fields in Appendix~\ref{tf:sec:axial}. Figure~\ref{fig7} isolates the corresponding meson energies,
\begin{equation}
\begin{aligned}
	E_\omega^{\rm cur}&=-\frac12\int d^3r\,
	g_\omega\boldsymbol\omega\cdot\mathbf j,\\
	E_\rho^{\rm cur}&=-\frac12\int d^3r\,
	g_\rho\boldsymbol\rho_3\cdot\mathbf j_3,
	\qquad \mathbf j_3=\mathbf j_n-\mathbf j_p .
\end{aligned}
\label{tf:eq:meson_energies}
\end{equation}
These are the spatial parts of the internal vector-field energies, with the Hartree factor $1/2$, while they are neither the direct energy of coupling to the external field nor the full energetic effect of time-reversal breaking. They vanish in the zero-field case and become negative as the field induces currents, providing attraction in this decomposition. Their values do not determine the sign of the local currents or the nuclear magnetization.

For $^{130}$Cd, the magnitude of the $\omega$-current energy increases overall with considerable local nonmonotonicity. At $10^{18}$~G, $E_\omega^{\rm cur}=(-1.69,\ -1.88)$~MeV and $E_\rho^{\rm cur}=(-0.39,\ -0.60)$~MeV. Together the two terms therefore contribute about $(-2.08,\ -2.47)$~MeV. This contribution is much smaller in magnitude than the individual scalar and time-like vector shifts. Their changes near $5.2$, $6.0$, and $9.0\times10^{17}$~G accompany changes in deformation. This is consistent with the common dependence on occupied spinors, while it does not establish that the current energy drives a particular shape switch.

The current response for $^{64}$Ge is especially nonmonotonic. The largest magnitudes of the $\omega$-current energy occur in the low-field region, having $1.28$~MeV at $5.6\times10^{16}$~G for NL3*, and $1.54$~MeV at $10^{16}$~G for DD-ME2. At $10^{18}$~G, the $\omega$ contributions are only $(-0.55,\ -0.63)$~MeV, while the $\rho$ contributions are $(-0.35,\ -0.46)$~MeV. It also demonstrates that a large current contribution does not necessarily imply a monotonic dependence on $B$, $\beta_2$, or the nuclear radius.

A finite isovector-current energy in this $N=Z$ nucleus is physically consistent, because equality of the integrated neutron and proton density numbers does not require $\mathbf j_n(\mathbf r)=\mathbf j_p(\mathbf r)$. The distinct orbital and magnetic-moment couplings of neutrons and protons allow for both sum and difference currents.

\begin{figure*}[t]
\centering
\includegraphics[width=\textwidth]{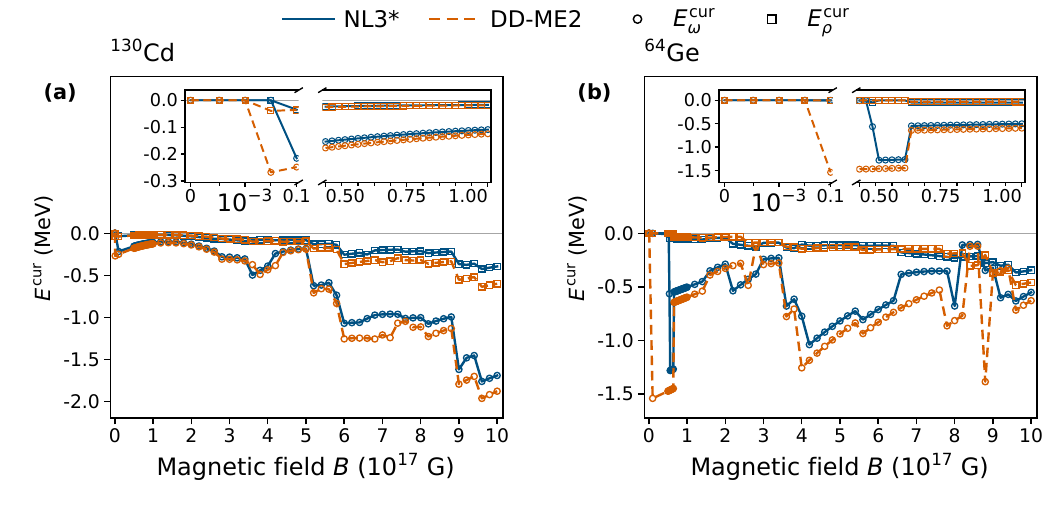}
\caption{Integrated $\omega$- and $\rho$-current energies for (a) $^{130}$Cd and (b) $^{64}$Ge. Open circles and squares denote the $E_\omega^{\rm cur}$ and $E_\rho^{\rm cur}$, respectively. Functional colors, line styles, and main/inset field-axis conventions as in Fig.~\ref{fig1}.}
\label{fig7}
\end{figure*}

\subsection{Spatial density distributions}
\label{subsec:densities}

\begin{figure*}[b]
\centering
\includegraphics[width=\textwidth]{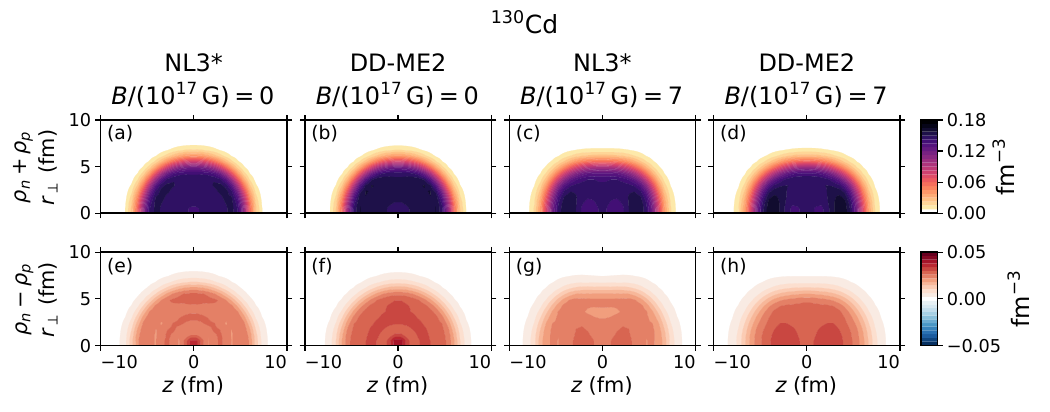}
\caption{Total, $\rho_n+\rho_p$, (upper row) and isovector, $\rho_n-\rho_p$, (lower row) number densities of $^{130}$Cd in the $(z,r_\perp)$ plane, with the magnetic field parallel to $z$. From left to right, the columns correspond to the NL3* and DD-ME2 functionals at $B=0$, followed by the same functionals at $B=7\times10^{17}$~G. Densities are in $\mathrm{fm}^{-3}$ and each row has a common color scale.}
\label{fig8}
\end{figure*}

\begin{figure*}[t]
\centering
\includegraphics[width=\textwidth]{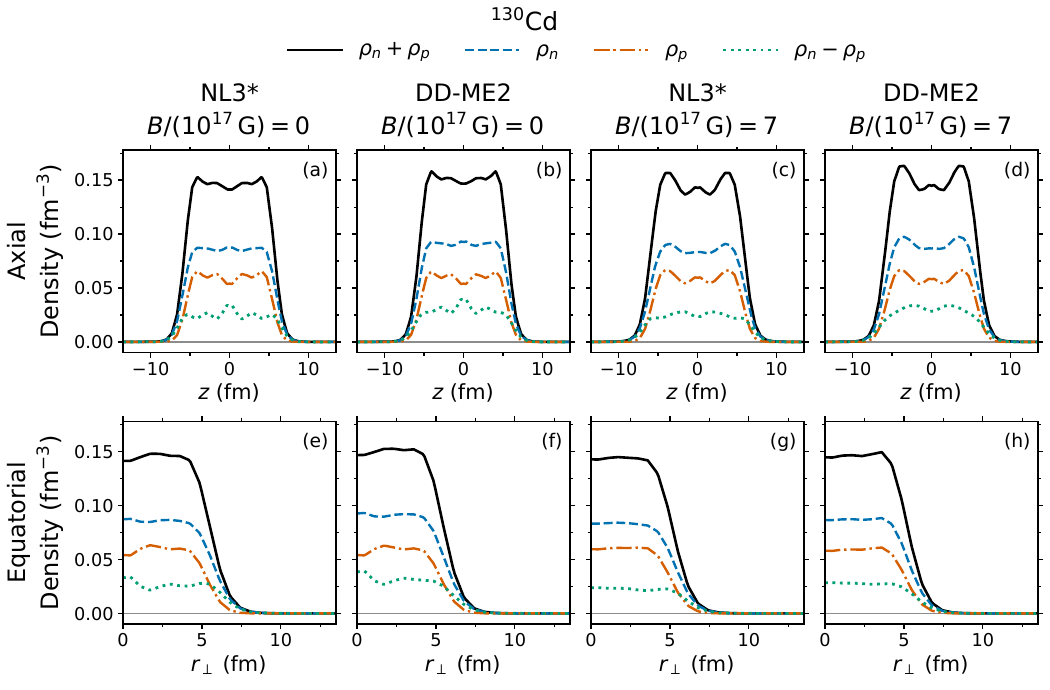}
\caption{Axial, $r_\perp=0$, (upper row) and equatorial, $z=0$, (lower row) number density profiles of $^{130}$Cd, with magnetic field values and column order as in Fig.~\ref{fig8}. Black solid, blue dashed, orange dash-dotted, and green dotted curves denote total, neutron, proton, and isovector densities, respectively.}
\label{fig9}
\end{figure*}

\begin{figure*}[t]
\centering
\includegraphics[width=\textwidth]{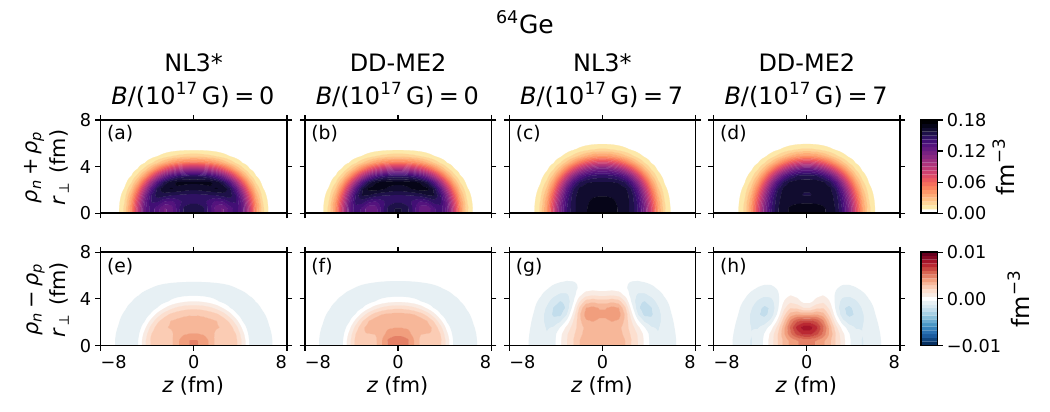}
\caption{Total, $\rho_n+\rho_p$, (upper row) and isovector, $\rho_n-\rho_p$, (lower row) number densities of $^{64}$Ge in the $(z,r_\perp)$ plane, with magnetic field values, column order and conventions as in Fig.~\ref{fig8}.}
\label{fig10}
\end{figure*}

\begin{figure*}[t]
\centering
\includegraphics[width=\textwidth]{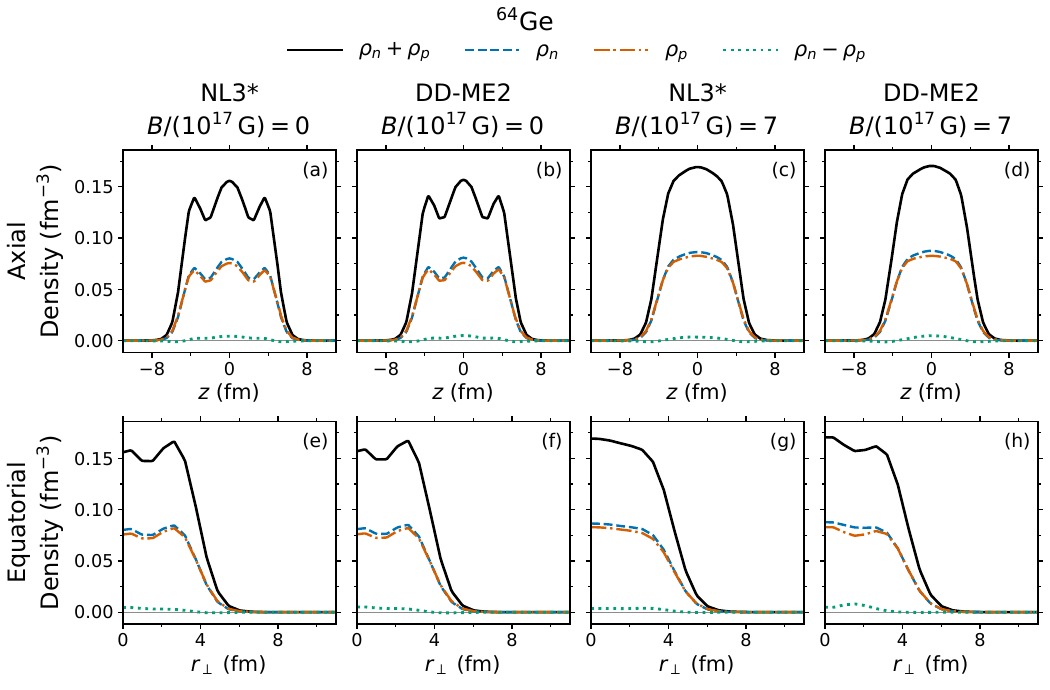}
\caption{Axial, $r_\perp=0$, (upper row) and equatorial, $z=0$, (lower row) number density profiles of $^{64}$Ge, with magnetic field values,  column order and conventions as in Fig.~\ref{fig9}.}
\label{fig11}
\end{figure*}

Figures~\ref{fig8}-\ref{fig11}  show, for $B=0$ and $7\times10^{17}$~G, the total number density $\rho_n+\rho_p=j^0$ and the isovector number density $\rho_n-\rho_p=j_3^0$, while the profiles are local axial and equatorial cuts. These number densities are distinct from the scalar source $\rho_s$ and, when currents are present, from the invariant $\rho_v$ entering DD-ME2. The selected field illustrates a prolate $^{130}$Cd configuration and a $^{64}$Ge configuration with a nearly vanishing total quadrupole deformation.

For $^{130}$Cd, the contours in Fig.~\ref{fig8} become more elongated parallel to the field, consistent with the change from $\beta_2\simeq0.04$ to $(0.196,\ 0.178)$. Figure~\ref{fig9} shows a broader axial total-density profile with more pronounced off-center maxima and a smaller transverse extent of the high-density region. This anisotropic redistribution accompanies increases of the total rms radius from $4.864$~fm to $5.017$~fm for NL3*, and from $4.808$~fm to $4.951$~fm for DD-ME2. The central total density in the profiles changes only slightly, from approximately $(0.141,\ 0.147)~\mathrm{fm}^{-3}$ to $(0.143,\ 0.145)~\mathrm{fm}^{-3}$.

Nevertheless, the central neutron excess decreases in both functionals, from approximately $(0.033,\ 0.038)~\mathrm{fm}^{-3}$ to $(0.024,\ 0.028)~\mathrm{fm}^{-3}$, while the neutron skin increases from $(0.395,\ 0.290)$~fm to $(0.482,\ 0.362)$~fm. The skin compares normalized second moments across the whole density, while the central isovector density measures a local neutron excess. A redistribution of neutrons toward larger distances can thicken the skin, while reducing that excess in the interior. Indeed, DD-ME2 has the larger central neutron excess, but the smaller neutron skin at both magnetic field values.

The response of $^{64}$Ge is qualitatively different. The initially prolate total-density contours in Fig.~\ref{fig10} become more nearly isotropic, and the oscillatory axial interior profile in Fig.~\ref{fig11} is replaced by a smoother central maximum. The central total density in the profiles rises from approximately $(0.156,\ 0.157)~\mathrm{fm}^{-3}$ to $(0.169,\ 0.170)~\mathrm{fm}^{-3}$, an increase of about $8.6\%$ for either functional. Nevertheless, the total rms radius increases slightly, from $3.857$~fm to $(3.875,\ 3.872)$~fm. The increased central density thus reflects redistribution within a changing shape, rather than uniform compression. A small total $\beta_2$ does not establish spherical symmetry. For example, at this field, the neutron and proton quadrupole deformations, using the NL3*, are $-0.0196$ and $+0.0085$, respectively, and their contributions to the total quadrupole moment partially cancel. Higher moments and the anisotropic isovector distribution also remain unconstrained by a vanishing total quadrupole moment.

For a nucleus with $N=Z$, $\int d^3r\,(\rho_n-\rho_p)=0$, so local positive and negative regions of the $^{64}$Ge isovector density plots represent redistribution between species. The nonzero-field plots exhibit peripheral lobes of proton excess and regions of neutron excess in the interior, consistent with the more extended proton distribution. Their detailed interior pattern differs between NL3* and DD-ME2, although the total profiles are similar. These features illustrate how a proton skin can coexist with a local neutron excess and why the isovector density contains information that is not captured by the total rms radius. The profiles could support the connection between spectral rearrangements, deformation, and species-dependent expansion, but assigning an individual density feature to a particular orbital would require its spatial contribution to be resolved.

\FloatBarrier

\section{Conclusions}
\label{sec:conclusions}

We have investigated the static response of $^{64}$Ge and $^{130}$Cd to uniform magnetic fields up to $10^{18}$~G, using the NL3* and DD-ME2 functionals.  Both functionals predict an overall decrease of the total energy and an increase of the rms radius at strong fields. These trends coexist with nonmonotonic changes in deformation and switches between prolate and oblate configurations. The expansion depends on the species: neutrons expand more in $^{130}$Cd, while protons expand more in $^{64}$Ge, increasing the respective neutron and proton skins overall. At $7\times10^{17}$~G, $^{130}$Cd develops a greater axial elongation than at zero field, while $^{64}$Ge has a more nearly isotropic total density and a higher central total density, although it has a slightly larger rms radius. The density plots thus show redistribution within the nucleus rather than uniform expansion or compression. In particular, neither a small total quadrupole moment nor the total radius alone determines the local neutron-proton distribution.

Microscopic analysis distinguishes the lifting of Kramers degeneracy from actual occupation changes. Full-spinor overlap matching, between zero field and $10^{17}$~G, shows that one descendant of the highest occupied zero-field proton doublet in $^{130}$Cd becomes unoccupied, while both descendants of lower-lying doublets for $^{64}$Ge remain occupied at that field. The neutron occupied-unoccupied spacing also evolves differently in the two nuclei. It decreases substantially in $^{130}$Cd, while in $^{64}$Ge it reopens at high fields after an initial decrease. Following states and occupations continuously across the configuration switches would clarify the microscopic origin of the abrupt bulk changes.

The energy decomposition clarifies how the individual contributions combine to produce the total-energy decrease. The scalar attraction and the time-like $\omega$ repulsion both weaken relative to zero field, producing large opposing shifts. Their positive net shift opposes the energy decrease. The attractive meson-current terms are much smaller than the dominant scalar and time-like vector changes. Their nonmonotonic field dependence, especially in $^{64}$Ge, emphasizes the configuration dependence of the time-odd response. NL3* and DD-ME2 agree on the principal strong-field trends, while retaining differences in the neutron skin of $^{130}$Cd, intermediate configurations, current energies, and local isovector densities. NL3* predicts a thicker neutron skin for $^{130}$Cd than DD-ME2 throughout the field range.

Extending these calculations to neighboring nuclei would provide field-dependent nucleon separation energies and reaction $Q$ values. For stellar weak rates, the relevant parent-daughter mass differences could be combined with nuclear transition strengths and the field-dependent lepton phase space. Reaction-network calculations incorporating these weak rates and the competing capture rates could then assess how strongly magnetic fields affect waiting-point lifetimes and abundance distributions.

\begin{acknowledgments}
The results presented in this work were obtained using the Aristotle University of Thessaloniki (AUTh) High Performance Computing Infrastructure and Resources. The authors also acknowledge GCP resources provided by the National Infrastructures for Research and Technology (GRNET) and funded by the European Union's Recovery and Resilience Facility within the project ``Towards the Quantum Cosmos: Deep Learning and Quantum Computation for Neutron Star Physics'' (ToQC).
\end{acknowledgments}

\appendix

\section{Axially symmetric formulation}
\label{tf:sec:axial}

We restrict the stationary solutions to axial symmetry about the magnetic-field direction and reflection symmetry across $z=0$, allowing axial deformation and retaining parity as a good quantum number. The radial spinor amplitudes are chosen real in this stationary solution class. In cylindrical coordinates $(r_\perp,\varphi,z)$, the scalar and time-like fields depend on $(r_\perp,z)$. Within this class, the spatial currents and vector potentials are azimuthal \cite{Arteagaetal2011}, namely, $\mathbf j=j_\varphi(r_\perp,z)\mathbf e_\varphi$ and $\boldsymbol\omega=\omega_\varphi(r_\perp,z)\mathbf e_\varphi$. The external potential is $\mathbf A^{\rm ext}=(Br_\perp/2)\mathbf e_\varphi$.

The angular-momentum projection $\Omega_k$ is the conserved eigenvalue of $J_z=-i\partial_\varphi+\Sigma_z/2$. A nucleon spinor of species $q=n,p$ can be written as
\begin{equation}
  \psi_{kq}(\mathbf r)=\frac{1}{\sqrt{2\pi}}
\begin{pmatrix}
    f^+_{kq}(r_\perp,z)e^{i(\Omega_k-1/2)\varphi}\\
    f^-_{kq}(r_\perp,z)e^{i(\Omega_k+1/2)\varphi}\\
    i g^+_{kq}(r_\perp,z)e^{i(\Omega_k-1/2)\varphi}\\
    i g^-_{kq}(r_\perp,z)e^{i(\Omega_k+1/2)\varphi}
\end{pmatrix}\chi_q,
\label{tf:eq:axial-spinor}
\end{equation}
where $\chi_q$ is the isospin eigenvector and $\Omega_k$ is half-integer. Define the two-component amplitudes $\boldsymbol f_{kq}=(f^+_{kq},f^-_{kq})^T$ and $\boldsymbol g_{kq}=(g^+_{kq},g^-_{kq})^T$, together with
\begin{equation}
\begin{aligned}
  \mathcal V_q&=g_\omega\omega^0+g_\rho t_q\rho_3^0
                   +e_p\delta_{qp}A^0+\Sigma_R^0,\\
  \mathcal W_q&=g_\omega\omega_\varphi+g_\rho t_q\rho_{3\varphi}
                   +e_p\delta_{qp}(A_\varphi+Br_\perp/2)
                   +\Sigma_{R\varphi},\\
  K_q&=\kappa_q\mu_N B,\qquad t_n=+1,\quad t_p=-1.
\end{aligned}
\label{tf:eq:axial-potentials}
\end{equation}
The subscript $\varphi$ denotes the component along the unit vector $\mathbf e_\varphi$, rather than a covariant coordinate component. With $s_z=\operatorname{diag}(1,-1)$, Eq.~\eqref{tf:eq:hamiltonian} reduces to
\begin{equation}
\begin{aligned}
  (\epsilon_{kq}-\mathcal V_q-M+K_qs_z)\boldsymbol f_{kq}
    &=\mathcal D_{\Omega_k q}\boldsymbol g_{kq},\\
  (\epsilon_{kq}-\mathcal V_q+M-K_qs_z)\boldsymbol g_{kq}
    &=-\mathcal D_{\Omega_k q}\boldsymbol f_{kq},
\end{aligned}
\label{tf:eq:axial-dirac}
\end{equation}
where
\begin{equation}
  \mathcal D_{\Omega q}=
\begin{pmatrix}
    \partial_z & \displaystyle\partial_{r_\perp}
          +\frac{\Omega+1/2}{r_\perp}-\mathcal W_q\\
    \displaystyle\partial_{r_\perp}
          -\frac{\Omega-1/2}{r_\perp}+\mathcal W_q & -\partial_z
\end{pmatrix}.
\label{tf:eq:axial-operator}
\end{equation}
Time-reversal symmetry is broken at finite $B$, so states with $+\Omega_k$ and $-\Omega_k$ must be treated independently, without imposing Kramers degeneracy.

For the scalar and time-like fields, we introduce
\begin{equation}
  \mathcal D_0=\partial_{r_\perp}^2+
               \frac1{r_\perp}\partial_{r_\perp}+\partial_z^2,
  \qquad \mathcal D_1=\mathcal D_0-\frac1{r_\perp^2}.
\label{tf:eq:axial-laplacians}
\end{equation}
The azimuthal vector Laplacian acts as $\nabla^2(w_\varphi\mathbf e_\varphi) =(\mathcal D_1 w_\varphi)\mathbf e_\varphi$. Since the azimuthal sources, including their local density-dependent vertices, are divergence-free, the field equations reduce to
\begin{equation}
\begin{aligned}
  (-\mathcal D_0+m_\sigma^2)\sigma
    &=-g_\sigma\rho_s-U_{\rm nl}'(\sigma),\\
  (-\mathcal D_0+m_\omega^2)\omega^0&=g_\omega j^0,\\
  (-\mathcal D_0+m_\rho^2)\rho_3^0&=g_\rho j_3^0,\\
  -\mathcal D_0 A^0&=e_pj_p^0,\\
  (-\mathcal D_1+m_\omega^2)\omega_\varphi&=g_\omega j_\varphi,\\
  (-\mathcal D_1+m_\rho^2)\rho_{3\varphi}&=g_\rho j_{3\varphi},\\
  -\mathcal D_1 A_\varphi&=e_pj_{p\varphi}.
\end{aligned}
\label{tf:eq:axial-fields}
\end{equation}
Coulomb gauge is used for the internal electromagnetic potential. The $1/r_\perp^2$ contribution in the spatial equations follows from the vector character of the fields and is absent from their time-like equations. Regularity on the symmetry axis requires vanishing radial derivatives of scalar and time-like fields, while azimuthal fields vanish linearly or faster as $r_\perp\to0$. The internal fields and bound nucleon wave functions vanish at infinity, but this condition is not imposed on the prescribed external potential.

At fixed $B$, the sources in Eq.~\eqref{tf:eq:sources} and the coupled equations \eqref{tf:eq:axial-dirac} and \eqref{tf:eq:axial-fields} determine the self-consistent solution. The time-like and spatial meson fields use the same vertices. For DD-ME2, the vertices and all components of the rearrangement self-energy are updated with the densities and currents.

\section{Nuclear energy and center-of-mass correction}
\label{tf:sec:energy}

Let $A=N+Z$. The free Dirac kinetic contribution, with the nucleon rest masses subtracted, is
\begin{equation}
  T_0=\sum_{k\in{\rm occ}}\int d^3r\,
  \psi_k^\dagger
  [\boldsymbol\alpha\cdot(-i\boldsymbol\nabla)+\beta m-m]
  \psi_k .
\label{tf:eq:free-energy}
\end{equation}
Using the internal field equations in the unpaired Hartree functional \cite{Arteagaetal2011} gives the energy relative to $Am$ as
\begin{equation}
\begin{aligned}
  E_{\rm H}={}&T_0+\frac12\int d^3r\,\mathcal I_{\rm int}
       +E_{\rm ext}\\
       &+\int d^3r\,[U_{\rm nl}-\tfrac12\sigma U_{\rm nl}'],\\
  \mathcal I_{\rm int}={}&g_\sigma\sigma\rho_s
          +g_\omega\omega_\mu j^\mu\\
       &+g_\rho\rho_{3\mu}j_3^\mu+e_p A_\mu j_p^\mu .
\end{aligned}
\label{tf:eq:hartree-energy}
\end{equation}
All vertices are evaluated locally. The internal vector contractions include the spatial-current terms, such as $-g_\omega\boldsymbol\omega\cdot\mathbf j$, with the same Hartree factor $1/2$ as their time-like counterparts.
The nonlinear scalar correction is
\begin{equation}
  U_{\rm nl}-\tfrac12\sigma U_{\rm nl}'
    =-\frac{g_2}{6}\sigma^3-\frac{g_3}{4}\sigma^4
\label{tf:eq:nonlinear-energy}
\end{equation}
for NL3* and vanishes for DD-ME2. Equation~\eqref{tf:eq:hartree-energy} uses $T_0$, rather than a sum of interacting single-particle eigenvalues.

The direct external-field contribution is
\begin{equation}
\begin{aligned}
  E_{\rm ext}={}&-e_p\int d^3r\,
           \mathbf A^{\rm ext}\cdot\mathbf j_p\\
       &-\mu_NB\sum_{k\in{\rm occ}}\int d^3r\,
          \psi_k^\dagger\beta\widehat\kappa\Sigma_z\psi_k .
\end{aligned}
\label{tf:eq:external-energy}
\end{equation}
Both prescribed-field couplings have full weight, and no external Maxwell energy is added. The DD-ME2 rearrangement self-energy enters the Hamiltonian, but is not an additional term in Eq.~\eqref{tf:eq:hartree-energy}. Reconstructing the energy from single-particle eigenvalues instead requires subtracting the one-body trace of the rearrangement self-energy, $\int d^3r\,\Sigma_{R\mu}j^\mu$, together with the internal-field double counting \cite{Fuchs1995}.

The reported total energy is $E_{\rm tot}=E_{\rm H}+E_{\rm cm}$. The center-of-mass correction uses the oscillator estimate \cite{Gambhir1990} for NL3* and the microscopic prescription \cite{PhysRevC.71.024312} for DD-ME2:
\begin{equation}
  E_{\rm cm}=
\begin{cases}
    -\tfrac34\hbar\omega, & \text{NL3*},\\
    -\langle\mathbf P^2\rangle/(2Am), & \text{DD-ME2},
\end{cases}
\label{tf:eq:cm-energy}
\end{equation}
where $\hbar\omega=41A^{-1/3}\,\mathrm{MeV}$ and $\mathbf P=\sum_{a=1}^A(-i\boldsymbol\nabla_a)$ is the total canonical momentum. These prescriptions are retained as approximations at finite $B$ in the centered symmetric gauge of Eq.~\eqref{tf:eq:gauge}. The DD-ME2 correction is evaluated after variation. Both $E_{\rm H}$ and $E_{\rm cm}$ are included when comparing configurations and their magnetic response.

\section{Quantum-number assignments for the single-particle spectra}
\label{app:additional_results}

Tables~\ref{tab:spe:Cd130:NL3STAR:proton} and \ref{tab:spe:Cd130:NL3STAR:neutron} specify the ranked proton and neutron energy levels displayed in Figs.~\ref{fig4} and \ref{fig5}. The colored overlap-matched states can lie outside this tabulated window. For the axial, parity-conserving solutions, the signed projection $\Omega$ of the total single-particle angular momentum on the symmetry axis and the parity $\pi$ label the symmetry sector. The integer $k$ orders the stored eigenstates within that sector at a given field.

The notation $\Omega^\pi[N\,n_z\,|\Lambda|]$ is supplemented by the spin projection $\Sigma$ of the largest-amplitude oscillator component of the upper Dirac spinor. Here $N$ and $n_z$ are the major and axial oscillator quantum numbers, respectively, and $\Lambda$ is the orbital angular-momentum projection of that component on the symmetry axis. With $n_r$ the radial oscillator quantum number, the component labels obey
\begin{equation}
\begin{gathered}
 N=n_z+2n_r+|\Lambda|,\\
 \Omega=\Lambda+\Sigma,\qquad \pi=(-1)^N,
\end{gathered}
\label{tf:eq:labels}
\end{equation}
where $\Lambda$ is signed and $\Sigma=\pm1/2$.
Although $\Omega$ and $\pi$ are conserved labels of the calculated state, $N$, $n_z$, $\Lambda$, and $\Sigma$ in the tables describe only its dominant upper component. They are not simultaneously exact quantum numbers of the complete mixed Dirac eigenstate, and the listed $\Sigma$ is not its spin expectation value. Changes of the dominant component can occur through mixing and do not by themselves identify a level crossing.

At $B=0$, the calculation stores the positive-$\Omega$ representative of each Kramers pair with multiplicity two. The correlated upper and lower signs in a table cell give the representative and its time-reversed partner, while $k$ retains the index of the stored representative. This pair degeneracy must be distinguished from the $2j+1$ degeneracy of a spherical shell. At finite field, signed-$\Omega$ states are stored individually and need not be degenerate. Ranks 1-4 select the four highest occupied stored levels in ascending energy, while ranks 5-7 select the three lowest unoccupied levels. In the present unpaired calculations, the selected levels have occupations zero or one. Both the ranks and the dominant-component assignments are evaluated independently at each field. Consequently, each row describes the spectrum at a single field. Figs.~\ref{fig4} and \ref{fig5} highlight the overlap-matched branches to compare states at the two field endpoints, $B=0$ and $10^{17}$~G, as discussed in Sec.~\ref{subsec:levels}.

\begin{table*}[t]
\caption{Proton single-particle assignments for $^{130}$Cd with NL3*. Each cell gives $\Omega^{\pi}[N\,n_z\,|\Lambda|]$, $\Sigma$ and $k$. Here $\Omega$ is the angular-momentum projection, $\pi$ is parity, and $k$ is the eigenstate index within the $\Omega^{\pi}$ sector. The oscillator indices and $\Sigma$ label the dominant upper-spinor component. Ranks 1-4 are the four highest occupied levels (rank 4 the highest), while ranks 5-7 are the three lowest unoccupied levels (rank 5 the lowest). At $B=0$, upper/lower signs denote Kramers pairs and $k$ labels the $+\Omega$ representative.}
\label{tab:spe:Cd130:NL3STAR:proton}
\centering
\begin{ruledtabular}
\begin{tabular}{c*{7}{c}}
$B$ & \multicolumn{4}{c}{Occupied} & \multicolumn{3}{c}{Unoccupied} \\
$(10^{17}\,\mathrm{G})$ & 1 & 2 & 3 & 4 & 5 & 6 & 7 \\
\colrule
0
 & $\pm\frac{1}{2}^{+}\,[4\,2\,0]$
 & $\pm\frac{3}{2}^{+}\,[4\,3\,1]$
 & $\pm\frac{5}{2}^{+}\,[4\,2\,2]$
 & $\pm\frac{7}{2}^{+}\,[4\,1\,3]$
 & $\pm\frac{9}{2}^{+}\,[4\,0\,4]$
 & $\pm\frac{1}{2}^{+}\,[4\,3\,1]$
 & $\pm\frac{3}{2}^{+}\,[4\,2\,2]$
\\
 & $\Sigma{=}\pm\frac{1}{2},k{=}5$
 & $\Sigma{=}\pm\frac{1}{2},k{=}3$
 & $\Sigma{=}\pm\frac{1}{2},k{=}2$
 & $\Sigma{=}\pm\frac{1}{2},k{=}1$
 & $\Sigma{=}\pm\frac{1}{2},k{=}1$
 & $\Sigma{=}\mp\frac{1}{2},k{=}6$
 & $\Sigma{=}\mp\frac{1}{2},k{=}4$
\\
1
 & $+\frac{1}{2}^{+}\,[4\,2\,0]$
 & $-\frac{1}{2}^{+}\,[4\,3\,1]$
 & $-\frac{3}{2}^{+}\,[4\,2\,2]$
 & $-\frac{5}{2}^{+}\,[4\,2\,2]$
 & $-\frac{7}{2}^{+}\,[4\,1\,3]$
 & $-\frac{9}{2}^{+}\,[4\,0\,4]$
 & $+\frac{7}{2}^{+}\,[4\,0\,4]$
\\
 & $\Sigma{=}+\frac{1}{2},k{=}5$
 & $\Sigma{=}+\frac{1}{2},k{=}5$
 & $\Sigma{=}+\frac{1}{2},k{=}3$
 & $\Sigma{=}-\frac{1}{2},k{=}2$
 & $\Sigma{=}-\frac{1}{2},k{=}1$
 & $\Sigma{=}-\frac{1}{2},k{=}1$
 & $\Sigma{=}-\frac{1}{2},k{=}2$
\\
2
 & $-\frac{1}{2}^{+}\,[4\,3\,1]$
 & $-\frac{3}{2}^{-}\,[3\,0\,1]$
 & $-\frac{3}{2}^{+}\,[4\,2\,2]$
 & $-\frac{5}{2}^{+}\,[4\,2\,2]$
 & $-\frac{7}{2}^{+}\,[4\,1\,3]$
 & $+\frac{7}{2}^{+}\,[4\,0\,4]$
 & $+\frac{5}{2}^{+}\,[4\,1\,3]$
\\
 & $\Sigma{=}+\frac{1}{2},k{=}5$
 & $\Sigma{=}-\frac{1}{2},k{=}4$
 & $\Sigma{=}+\frac{1}{2},k{=}3$
 & $\Sigma{=}-\frac{1}{2},k{=}2$
 & $\Sigma{=}-\frac{1}{2},k{=}1$
 & $\Sigma{=}-\frac{1}{2},k{=}2$
 & $\Sigma{=}-\frac{1}{2},k{=}3$
\\
3
 & $-\frac{1}{2}^{-}\,[3\,3\,0]$
 & $-\frac{3}{2}^{-}\,[3\,0\,1]$
 & $-\frac{3}{2}^{+}\,[4\,2\,2]$
 & $+\frac{11}{2}^{-}\,[5\,0\,5]$
 & $+\frac{7}{2}^{+}\,[4\,0\,4]$
 & $+\frac{5}{2}^{+}\,[4\,0\,2]$
 & $+\frac{9}{2}^{-}\,[5\,1\,4]$
\\
 & $\Sigma{=}-\frac{1}{2},k{=}6$
 & $\Sigma{=}-\frac{1}{2},k{=}4$
 & $\Sigma{=}+\frac{1}{2},k{=}3$
 & $\Sigma{=}+\frac{1}{2},k{=}1$
 & $\Sigma{=}-\frac{1}{2},k{=}2$
 & $\Sigma{=}+\frac{1}{2},k{=}3$
 & $\Sigma{=}+\frac{1}{2},k{=}1$
\\
4
 & $-\frac{1}{2}^{+}\,[4\,1\,1]$
 & $+\frac{9}{2}^{-}\,[5\,1\,4]$
 & $+\frac{7}{2}^{+}\,[4\,0\,4]$
 & $+\frac{5}{2}^{+}\,[4\,0\,2]$
 & $-\frac{1}{2}^{-}\,[3\,3\,0]$
 & $-\frac{3}{2}^{+}\,[4\,2\,2]$
 & $-\frac{3}{2}^{-}\,[3\,0\,1]$
\\
 & $\Sigma{=}+\frac{1}{2},k{=}5$
 & $\Sigma{=}+\frac{1}{2},k{=}1$
 & $\Sigma{=}-\frac{1}{2},k{=}2$
 & $\Sigma{=}+\frac{1}{2},k{=}3$
 & $\Sigma{=}-\frac{1}{2},k{=}6$
 & $\Sigma{=}+\frac{1}{2},k{=}3$
 & $\Sigma{=}-\frac{1}{2},k{=}4$
\\
5
 & $-\frac{7}{2}^{-}\,[3\,0\,3]$
 & $-\frac{5}{2}^{-}\,[3\,1\,2]$
 & $+\frac{1}{2}^{-}\,[3\,0\,1]$
 & $-\frac{1}{2}^{+}\,[4\,1\,1]$
 & $+\frac{7}{2}^{-}\,[5\,2\,3]$
 & $-\frac{3}{2}^{+}\,[4\,2\,2]$
 & $+\frac{5}{2}^{+}\,[4\,1\,3]$
\\
 & $\Sigma{=}-\frac{1}{2},k{=}1$
 & $\Sigma{=}-\frac{1}{2},k{=}2$
 & $\Sigma{=}-\frac{1}{2},k{=}6$
 & $\Sigma{=}+\frac{1}{2},k{=}5$
 & $\Sigma{=}+\frac{1}{2},k{=}2$
 & $\Sigma{=}+\frac{1}{2},k{=}3$
 & $\Sigma{=}-\frac{1}{2},k{=}4$
\\
6
 & $+\frac{3}{2}^{+}\,[4\,1\,1]$
 & $+\frac{5}{2}^{+}\,[4\,0\,2]$
 & $+\frac{5}{2}^{-}\,[5\,3\,2]$
 & $+\frac{1}{2}^{+}\,[4\,4\,0]$
 & $-\frac{1}{2}^{-}\,[3\,1\,0]$
 & $-\frac{5}{2}^{-}\,[3\,1\,2]$
 & $-\frac{3}{2}^{+}\,[4\,2\,2]$
\\
 & $\Sigma{=}+\frac{1}{2},k{=}4$
 & $\Sigma{=}+\frac{1}{2},k{=}3$
 & $\Sigma{=}+\frac{1}{2},k{=}3$
 & $\Sigma{=}+\frac{1}{2},k{=}6$
 & $\Sigma{=}-\frac{1}{2},k{=}6$
 & $\Sigma{=}-\frac{1}{2},k{=}2$
 & $\Sigma{=}+\frac{1}{2},k{=}3$
\\
7
 & $+\frac{1}{2}^{+}\,[4\,4\,0]$
 & $+\frac{5}{2}^{+}\,[4\,0\,2]$
 & $-\frac{5}{2}^{-}\,[3\,0\,3]$
 & $-\frac{3}{2}^{-}\,[3\,2\,1]$
 & $+\frac{3}{2}^{-}\,[5\,4\,1]$
 & $-\frac{3}{2}^{+}\,[4\,2\,2]$
 & $-\frac{1}{2}^{-}\,[3\,1\,0]$
\\
 & $\Sigma{=}+\frac{1}{2},k{=}6$
 & $\Sigma{=}+\frac{1}{2},k{=}3$
 & $\Sigma{=}+\frac{1}{2},k{=}1$
 & $\Sigma{=}-\frac{1}{2},k{=}3$
 & $\Sigma{=}+\frac{1}{2},k{=}5$
 & $\Sigma{=}+\frac{1}{2},k{=}3$
 & $\Sigma{=}-\frac{1}{2},k{=}6$
\\
8
 & $+\frac{5}{2}^{+}\,[4\,0\,2]$
 & $+\frac{1}{2}^{+}\,[4\,2\,0]$
 & $+\frac{3}{2}^{-}\,[5\,4\,1]$
 & $-\frac{5}{2}^{-}\,[3\,0\,3]$
 & $-\frac{3}{2}^{-}\,[3\,2\,1]$
 & $+\frac{11}{2}^{+}\,[6\,1\,5]$
 & $+\frac{13}{2}^{+}\,[6\,0\,6]$
\\
 & $\Sigma{=}+\frac{1}{2},k{=}3$
 & $\Sigma{=}+\frac{1}{2},k{=}6$
 & $\Sigma{=}+\frac{1}{2},k{=}5$
 & $\Sigma{=}+\frac{1}{2},k{=}1$
 & $\Sigma{=}-\frac{1}{2},k{=}3$
 & $\Sigma{=}+\frac{1}{2},k{=}1$
 & $\Sigma{=}+\frac{1}{2},k{=}1$
\\
9
 & $-\frac{1}{2}^{-}\,[3\,3\,0]$
 & $+\frac{13}{2}^{+}\,[6\,0\,6]$
 & $+\frac{3}{2}^{-}\,[5\,4\,1]$
 & $+\frac{11}{2}^{+}\,[6\,1\,5]$
 & $-\frac{5}{2}^{+}\,[2\,0\,2]$
 & $+\frac{9}{2}^{+}\,[6\,2\,4]$
 & $-\frac{5}{2}^{-}\,[3\,0\,3]$
\\
 & $\Sigma{=}-\frac{1}{2},k{=}5$
 & $\Sigma{=}+\frac{1}{2},k{=}1$
 & $\Sigma{=}+\frac{1}{2},k{=}5$
 & $\Sigma{=}+\frac{1}{2},k{=}1$
 & $\Sigma{=}-\frac{1}{2},k{=}1$
 & $\Sigma{=}+\frac{1}{2},k{=}2$
 & $\Sigma{=}+\frac{1}{2},k{=}1$
\\
10
 & $+\frac{11}{2}^{+}\,[6\,1\,5]$
 & $-\frac{1}{2}^{+}\,[2\,0\,0]$
 & $+\frac{3}{2}^{-}\,[5\,4\,1]$
 & $+\frac{9}{2}^{+}\,[6\,2\,4]$
 & $-\frac{1}{2}^{-}\,[3\,3\,0]$
 & $+\frac{7}{2}^{+}\,[6\,3\,3]$
 & $-\frac{5}{2}^{-}\,[3\,0\,3]$
\\
 & $\Sigma{=}+\frac{1}{2},k{=}1$
 & $\Sigma{=}-\frac{1}{2},k{=}5$
 & $\Sigma{=}+\frac{1}{2},k{=}5$
 & $\Sigma{=}+\frac{1}{2},k{=}2$
 & $\Sigma{=}-\frac{1}{2},k{=}5$
 & $\Sigma{=}+\frac{1}{2},k{=}2$
 & $\Sigma{=}+\frac{1}{2},k{=}1$
\end{tabular}
\end{ruledtabular}
\end{table*}

\begin{table*}[t]
\caption{Neutron single-particle assignments for $^{130}$Cd with NL3*, with notations and conventions as in Table~\ref{tab:spe:Cd130:NL3STAR:proton}.}
\label{tab:spe:Cd130:NL3STAR:neutron}
\centering
\begin{ruledtabular}
\begin{tabular}{c*{7}{c}}
$B$ & \multicolumn{4}{c}{Occupied} & \multicolumn{3}{c}{Unoccupied} \\
$(10^{17}\,\mathrm{G})$ & 1 & 2 & 3 & 4 & 5 & 6 & 7 \\
\colrule
0
 & $\pm\frac{5}{2}^{-}\,[5\,3\,2]$
 & $\pm\frac{7}{2}^{-}\,[5\,2\,3]$
 & $\pm\frac{9}{2}^{-}\,[5\,1\,4]$
 & $\pm\frac{11}{2}^{-}\,[5\,0\,5]$
 & $\pm\frac{1}{2}^{-}\,[5\,5\,0]$
 & $\pm\frac{3}{2}^{-}\,[5\,4\,1]$
 & $\pm\frac{5}{2}^{-}\,[5\,1\,2]$
\\
 & $\Sigma{=}\pm\frac{1}{2},k{=}3$
 & $\Sigma{=}\pm\frac{1}{2},k{=}2$
 & $\Sigma{=}\pm\frac{1}{2},k{=}1$
 & $\Sigma{=}\pm\frac{1}{2},k{=}1$
 & $\Sigma{=}\pm\frac{1}{2},k{=}8$
 & $\Sigma{=}\pm\frac{1}{2},k{=}6$
 & $\Sigma{=}\pm\frac{1}{2},k{=}4$
\\
1
 & $-\frac{11}{2}^{-}\,[5\,0\,5]$
 & $+\frac{7}{2}^{-}\,[5\,2\,3]$
 & $+\frac{9}{2}^{-}\,[5\,1\,4]$
 & $+\frac{11}{2}^{-}\,[5\,0\,5]$
 & $-\frac{3}{2}^{-}\,[5\,4\,1]$
 & $-\frac{1}{2}^{-}\,[5\,5\,0]$
 & $-\frac{5}{2}^{-}\,[5\,1\,2]$
\\
 & $\Sigma{=}-\frac{1}{2},k{=}1$
 & $\Sigma{=}+\frac{1}{2},k{=}2$
 & $\Sigma{=}+\frac{1}{2},k{=}1$
 & $\Sigma{=}+\frac{1}{2},k{=}1$
 & $\Sigma{=}-\frac{1}{2},k{=}6$
 & $\Sigma{=}-\frac{1}{2},k{=}8$
 & $\Sigma{=}-\frac{1}{2},k{=}4$
\\
2
 & $+\frac{5}{2}^{-}\,[5\,3\,2]$
 & $+\frac{7}{2}^{-}\,[5\,2\,3]$
 & $+\frac{9}{2}^{-}\,[5\,1\,4]$
 & $+\frac{11}{2}^{-}\,[5\,0\,5]$
 & $-\frac{3}{2}^{-}\,[5\,4\,1]$
 & $-\frac{5}{2}^{-}\,[5\,1\,2]$
 & $-\frac{1}{2}^{-}\,[5\,5\,0]$
\\
 & $\Sigma{=}+\frac{1}{2},k{=}3$
 & $\Sigma{=}+\frac{1}{2},k{=}2$
 & $\Sigma{=}+\frac{1}{2},k{=}1$
 & $\Sigma{=}+\frac{1}{2},k{=}1$
 & $\Sigma{=}-\frac{1}{2},k{=}6$
 & $\Sigma{=}-\frac{1}{2},k{=}4$
 & $\Sigma{=}-\frac{1}{2},k{=}8$
\\
3
 & $+\frac{5}{2}^{-}\,[5\,2\,3]$
 & $+\frac{7}{2}^{-}\,[5\,2\,3]$
 & $+\frac{9}{2}^{-}\,[5\,1\,4]$
 & $+\frac{11}{2}^{-}\,[5\,0\,5]$
 & $-\frac{7}{2}^{-}\,[5\,0\,3]$
 & $-\frac{3}{2}^{-}\,[5\,4\,1]$
 & $-\frac{5}{2}^{-}\,[5\,1\,2]$
\\
 & $\Sigma{=}-\frac{1}{2},k{=}3$
 & $\Sigma{=}+\frac{1}{2},k{=}2$
 & $\Sigma{=}+\frac{1}{2},k{=}1$
 & $\Sigma{=}+\frac{1}{2},k{=}1$
 & $\Sigma{=}-\frac{1}{2},k{=}3$
 & $\Sigma{=}-\frac{1}{2},k{=}6$
 & $\Sigma{=}-\frac{1}{2},k{=}4$
\\
4
 & $+\frac{5}{2}^{-}\,[5\,2\,3]$
 & $+\frac{11}{2}^{-}\,[5\,0\,5]$
 & $-\frac{1}{2}^{+}\,[4\,3\,1]$
 & $+\frac{1}{2}^{+}\,[4\,4\,0]$
 & $-\frac{7}{2}^{-}\,[5\,0\,3]$
 & $-\frac{3}{2}^{-}\,[5\,0\,1]$
 & $+\frac{5}{2}^{-}\,[5\,0\,3]$
\\
 & $\Sigma{=}-\frac{1}{2},k{=}3$
 & $\Sigma{=}+\frac{1}{2},k{=}1$
 & $\Sigma{=}+\frac{1}{2},k{=}9$
 & $\Sigma{=}+\frac{1}{2},k{=}9$
 & $\Sigma{=}-\frac{1}{2},k{=}3$
 & $\Sigma{=}-\frac{1}{2},k{=}6$
 & $\Sigma{=}-\frac{1}{2},k{=}4$
\\
5
 & $+\frac{11}{2}^{-}\,[5\,0\,5]$
 & $+\frac{3}{2}^{+}\,[4\,3\,1]$
 & $-\frac{7}{2}^{-}\,[5\,0\,3]$
 & $-\frac{1}{2}^{+}\,[4\,3\,1]$
 & $-\frac{3}{2}^{-}\,[5\,0\,1]$
 & $+\frac{1}{2}^{-}\,[5\,0\,1]$
 & $+\frac{5}{2}^{-}\,[5\,0\,3]$
\\
 & $\Sigma{=}+\frac{1}{2},k{=}1$
 & $\Sigma{=}+\frac{1}{2},k{=}6$
 & $\Sigma{=}-\frac{1}{2},k{=}3$
 & $\Sigma{=}+\frac{1}{2},k{=}9$
 & $\Sigma{=}-\frac{1}{2},k{=}6$
 & $\Sigma{=}-\frac{1}{2},k{=}8$
 & $\Sigma{=}-\frac{1}{2},k{=}4$
\\
6
 & $-\frac{7}{2}^{+}\,[4\,0\,4]$
 & $+\frac{7}{2}^{-}\,[5\,1\,4]$
 & $+\frac{5}{2}^{+}\,[4\,0\,2]$
 & $-\frac{5}{2}^{-}\,[5\,1\,2]$
 & $-\frac{1}{2}^{-}\,[5\,1\,0]$
 & $+\frac{3}{2}^{-}\,[5\,1\,2]$
 & $-\frac{7}{2}^{-}\,[5\,0\,3]$
\\
 & $\Sigma{=}+\frac{1}{2},k{=}2$
 & $\Sigma{=}-\frac{1}{2},k{=}2$
 & $\Sigma{=}+\frac{1}{2},k{=}4$
 & $\Sigma{=}-\frac{1}{2},k{=}4$
 & $\Sigma{=}-\frac{1}{2},k{=}9$
 & $\Sigma{=}-\frac{1}{2},k{=}6$
 & $\Sigma{=}-\frac{1}{2},k{=}3$
\\
7
 & $-\frac{5}{2}^{-}\,[5\,1\,2]$
 & $-\frac{1}{2}^{+}\,[4\,1\,1]$
 & $-\frac{1}{2}^{-}\,[5\,1\,0]$
 & $+\frac{3}{2}^{-}\,[5\,1\,2]$
 & $-\frac{7}{2}^{+}\,[4\,0\,4]$
 & $+\frac{5}{2}^{+}\,[4\,0\,2]$
 & $-\frac{7}{2}^{-}\,[5\,0\,3]$
\\
 & $\Sigma{=}-\frac{1}{2},k{=}4$
 & $\Sigma{=}+\frac{1}{2},k{=}9$
 & $\Sigma{=}-\frac{1}{2},k{=}9$
 & $\Sigma{=}-\frac{1}{2},k{=}6$
 & $\Sigma{=}+\frac{1}{2},k{=}2$
 & $\Sigma{=}+\frac{1}{2},k{=}4$
 & $\Sigma{=}-\frac{1}{2},k{=}3$
\\
8
 & $+\frac{3}{2}^{+}\,[4\,1\,1]$
 & $-\frac{1}{2}^{-}\,[5\,1\,0]$
 & $+\frac{3}{2}^{-}\,[5\,1\,2]$
 & $-\frac{1}{2}^{+}\,[4\,1\,1]$
 & $-\frac{7}{2}^{-}\,[5\,0\,3]$
 & $-\frac{3}{2}^{-}\,[5\,0\,1]$
 & $+\frac{1}{2}^{-}\,[5\,0\,1]$
\\
 & $\Sigma{=}+\frac{1}{2},k{=}6$
 & $\Sigma{=}-\frac{1}{2},k{=}9$
 & $\Sigma{=}-\frac{1}{2},k{=}6$
 & $\Sigma{=}+\frac{1}{2},k{=}9$
 & $\Sigma{=}-\frac{1}{2},k{=}3$
 & $\Sigma{=}-\frac{1}{2},k{=}7$
 & $\Sigma{=}-\frac{1}{2},k{=}9$
\\
9
 & $+\frac{3}{2}^{-}\,[5\,1\,2]$
 & $-\frac{5}{2}^{+}\,[4\,1\,3]$
 & $+\frac{3}{2}^{+}\,[4\,1\,1]$
 & $-\frac{7}{2}^{-}\,[5\,0\,3]$
 & $-\frac{3}{2}^{-}\,[5\,0\,1]$
 & $+\frac{1}{2}^{-}\,[5\,0\,1]$
 & $-\frac{1}{2}^{+}\,[4\,1\,1]$
\\
 & $\Sigma{=}-\frac{1}{2},k{=}6$
 & $\Sigma{=}+\frac{1}{2},k{=}4$
 & $\Sigma{=}+\frac{1}{2},k{=}6$
 & $\Sigma{=}-\frac{1}{2},k{=}3$
 & $\Sigma{=}-\frac{1}{2},k{=}7$
 & $\Sigma{=}-\frac{1}{2},k{=}9$
 & $\Sigma{=}+\frac{1}{2},k{=}9$
\\
10
 & $+\frac{1}{2}^{+}\,[4\,2\,0]$
 & $-\frac{7}{2}^{-}\,[5\,0\,3]$
 & $-\frac{5}{2}^{+}\,[4\,1\,3]$
 & $-\frac{3}{2}^{-}\,[5\,0\,1]$
 & $+\frac{1}{2}^{-}\,[5\,0\,1]$
 & $+\frac{5}{2}^{-}\,[5\,0\,3]$
 & $+\frac{9}{2}^{-}\,[5\,0\,5]$
\\
 & $\Sigma{=}+\frac{1}{2},k{=}8$
 & $\Sigma{=}-\frac{1}{2},k{=}3$
 & $\Sigma{=}+\frac{1}{2},k{=}4$
 & $\Sigma{=}-\frac{1}{2},k{=}7$
 & $\Sigma{=}-\frac{1}{2},k{=}9$
 & $\Sigma{=}-\frac{1}{2},k{=}4$
 & $\Sigma{=}-\frac{1}{2},k{=}1$
\end{tabular}
\end{ruledtabular}
\end{table*}

\FloatBarrier

\bibliography{references_new}

\end{document}